\documentclass[superscriptaddress,twocolumn,aps,prc,showpacs]{revtex4-1}

\usepackage{amsmath}
\usepackage{amssymb}
\usepackage{graphicx}
\usepackage{mathrsfs}
\usepackage{color}
\newcommand{\bvec}[1]{\mbox{\boldmath $#1$}}

\usepackage[dvipdfm,bookmarks=true,colorlinks,%
            citecolor=blue,linkcolor=blue,hypertex, %
            breaklinks=true]{hyperref}

\begin{document}

\title{Three-dimensional mesh calculations for covariant density functional theory}

\author{Y. Tanimura}
\altaffiliation[Present address: ]{Institut de Physique Nucl\'eaire, IN2P3-CNRS, Universit\'e Paris-Sud, F-91406 Orsay Cedex, France}
\affiliation{Department of Physics, Tohoku University, Sendai 980-8578, Japan}
\author{K. Hagino}
\affiliation{Department of Physics, Tohoku University, Sendai 980-8578, Japan}
\affiliation{Research Center for Electron Photon Science, Tohoku University, 1-2-1 Mikamine, Sendai 982-0826, Japan}
\author{H. Z. Liang}
\affiliation{RIKEN Nishina Center, Wako 351-0198, Japan}

\begin{abstract}
In contrast to the non-relativistic approaches,
three-dimensional (3D) mesh calculations for the {\it relativistic} density functional theory have not been realized because of the challenges of variational collapse and fermion doubling.
We overcome these difficulties by developing a novel method based on the ideas of Wilson fermion as well as the variational principle for the inverse Hamiltonian.
We demonstrate the applicability of this method by applying it to $^{16}$O, $^{24}$Mg, and $^{28}$Si nuclei,
providing
detailed explanation on the formalism
and verification of numerical implementation.
\end{abstract}

\pacs{21.60.Jz, 21.10.Dr, 21.10.Gv, 03.65.Pm}

\maketitle

\section{Introduction}\label{sec:intro}

The self-consistent mean-field theory with phenomenological interactions has been successfully employed in nuclear physics in
the past decades \cite{Bender2003}.
This theory is intimately related to the density functional theory (DFT) originally developed for many electron systems \cite{Hohenberg1964, Kohn1965, Engel2011}.
The DFT is based on the Hohenberg-Kohn theorem \cite{Hohenberg1964, Engel2011} for the existence of a universal energy
density functional for many-body systems, which completely contains the many-body correlations in principle.
In nuclear systems, a large part of the many-body correlations is taken into account through the parameters of phenomenological interactions, which are determined by fitting to a selected set of
experimental data.
The resultant Hartree-Fock equation can thus be regarded as the Kohn-Sham equation \cite{Kohn1965, Engel2011} in the DFT, which already contains the correlation effects even though it appears an equation for
non-interacting systems.
Notice that, starting from an energy density functional, one does not have to rely on a specific nucleon-nucleon interaction, which is an important feature in considering the density dependence of the energy functional.
Energy functionals of Skyrme \cite{Skyrme1959} and Gogny \cite{Decharge1980} types have been developed for the non-relativistic nuclear many-body calculations.
The relativistic variant of DFT, referred to as the covariant density functional theory (CDFT), has also been developed
\cite{Ring1996, Lalazissis2004, Vretenar2005, Meng2006, Niksic2011, Engel2011}.

The nuclear DFT has several advantages over other theoretical methods.
Firstly, since it reduces a problem of an interacting system to a problem of a non-interacting system, the numerical cost increases moderately with the number of particles in the system.
Because of this, the DFT is the only method at present which is able to describe atomic nuclei in the whole nuclear chart at a reasonable computational cost.
Secondly, formulated in the body-fixed frame, the DFT provides an intuitive view of nuclear deformation as a spontaneous breaking of symmetries.
As a matter of fact, a variety of deformation degrees of freedom are expected to play an important role in nuclear phenomena.
For instance, it has been pointed out that the non-axial and reflection-asymmetric deformations play important roles in the fission barriers as well as the fission paths \cite{Lu2012, Lu2014MDC}.
Also, the cluster states are expected to emerge in the excited states of $sd$-shell nuclei \cite{Beck2010, *Beck2012, *Beck2014, Kanada-Enyo2012, Ebran2012, Ebran2013, Ebran2014, Girod2013}, and the exotic octupole and hexadecapole deformations are expected in several regions in the nuclear chart \cite{Spear1990, Hamamoto1991, Li1994, Butler1996, Takami1998, Matsuo1999, Yamagami2001, Dudek2002, Dudek2006, Dudek2014, Robledo2011, Zhao2012, Li2013}.

In order to describe various nuclear deformations flexibly and efficiently with DFT, a coordinate-space representation, in which the full real space is discretized into three-dimensional (3D) lattice, is most suitable.
Notice that with the basis expansion method, in which wave functions are expanded on a finite basis set, the solutions strongly reflect the properties of the basis used, and the method
is less powerful as compared to the coordinate-space representation.
For the nuclear DFT calculations with the non-relativistic Skyrme functional in the 3D coordinate-space representation, the so-called imaginary time method has been successfully developed \cite{Davies1980, Bonche2005}.
In this method, starting with arbitrary single-particle wave functions, one obtains self-consistent solutions after evolving the wave functions along the imaginary time axis.
It has been extensively applied to calculations not only for the ground state \cite{Tajima1996, Tajima2001} but also for excited states of atomic nuclei.
The latter includes exotic excitation modes in neutron-rich nuclei \cite{Inakura2011} and exotic structure in high-spin states \cite{Yamagami2000, Ichikawa2011, Ichikawa2012}.
The method has also been applied to deformation of $\Lambda$ hypernuclei \cite{Win2011} as well as to deformed Hartree-Fock-Bogoliubov calculations with the two-basis method \cite{Terasaki1996, Yamagami2001}.

In contrast to the non-relativistic case with Skyrme functional, the CDFT calculations for deformed nuclei have been performed only with the basis expansion method, see, e.g., Refs.~\cite{Zhou2010, Lu2011, Lu2012, Lu2014MDC, Lu2014Lambda, Zhao2012, Li2012, Chen2012}.
That is,
the CDFT calculations in the 3D coordinate space have not been realized yet.
There are two main reasons for this,
which are associated to the following essential challenges.
The first challenge is the existence of the Dirac sea,
in which the negative-energy spectrum is not bound from the bottom,
in a relativistic Hamiltonian.
In most of relativistic treatments, the densities are constructed with the occupied states in the Fermi sea only, explicitly excluding by hand all the states in the Dirac sea. This is referred to as
the no-sea approximation, and is
equivalent to neglecting the vacuum polarization effect \cite{Ring1996}.
In the no-sea approximation, the ground-state solution always corresponds to a saddle point on the energy surface, while in the non-relativistic case the ground state corresponds to the absolute minimum.
It prevents a direct application of the imaginary time evolution for the coordinate-space calculations to the relativistic systems.
That is, an iterative solution inevitably dives into the Dirac sea during the imaginary time evolution, leading to a divergence of the solution \cite{Zhang2009CPC, Zhang2009CPL, Zhang2010, Hagino2010}.
This problem is known as {\it variational collapse}.
The second challenge in the relativistic calculations is the {\it fermion doubling} problem \cite{Wilson1977, Kogut1983, Lee1986, Poschl1996, Busic1999, Fillion-Gourdeau2012, Hernandez2012}.
When one tries to solve a Dirac equation on discretized lattice in the real space, one tends to obtain spurious solutions with rapid oscillations as a function of coordinate, despite that the corresponding expectation value of energy is small.
This problem arises as long as the first derivative in a Dirac Hamiltonian is approximated by a finite difference, and has been in fact well known in the field of lattice quantum chromodynamics (QCD) \cite{Wilson1977, Kogut1983}.

In this paper, we propose a novel and practical method for the CDFT calculations on 3D lattice, overcoming the problems of the variational collapse and the fermion doubling.
For the first problem, we employ the inverse Hamiltonian method, which have been developed in the previous publication \cite{Hagino2010}.
While in Ref.~\cite{Hagino2010} we treated only the radial coordinate for a spherical relativistic Hamiltonian, in this paper
we extend it to a 3D space.
For the second problem, we extend the method of Wilson fermions, which has been widely adopted in the lattice QCD calculations \cite{Wilson1977, Kogut1983}.
With these strategies, we realize for the first time the relativistic calculations on 3D lattice without assuming any spatial symmetry.

The paper is organized as follows. In Sec.~\ref{sec:model}, we will briefly recall the relativistic point-coupling model for the CDFT calculations.
In Secs.~\ref{sec:VC} and \ref{sec:FD}, the two major
difficulties, that is,
the variational collapse and the fermion doubling, in the coordinate-space CDFT calculations will be explained in detail.
We will provide and discuss the strategy to resolve each of these difficulties.
In Sec. \ref{sec:SC}, we will numerically apply our method firstly to a spherical nucleus $^{16}$O and check the validity of the calculations.
We will then apply the method to deformed nuclei $^{24}$Mg and $^{28}$Si and discuss its applicability.
We will summarize the paper in Sec. \ref{sec:sum}.

\section{Relativistic point-coupling model}\label{sec:model}

The relativistic variant of DFT, referred to as CDFT \cite{Ring1996, Lalazissis2004, Vretenar2005, Meng2006, Niksic2011, Engel2011}, has been employed as widely as the non-relativistic DFT in the studies of nuclear structure.
A covariant energy density functional is obtained from a Lorentz invariant effective Lagrangian, which describes an effective interaction among nucleons via meson exchanges.
The relativistic treatment of nuclei has several characteristic features mainly due to the Lorentz invariance, which is the most important underlying symmetry of QCD.
The CDFT has attracted much attention during the past decades because of its success in many aspects in nuclear physics.
For instance, the saturation mechanism of nuclear matter \cite{Walecka1974, Serot1986}, the large spin-orbit splittings which yield the shell structure of nuclei \cite{Horowitz1981, Serot1986}, and the origin of the pseudospin symmetry \cite{Ginocchio1997, Ginocchio2005, Liang2011} are consistently understood with CDFT as the consequences of a delicate balance between the large attractive scalar and repulsive vector fields in nuclear medium.
The time-odd components in the mean field, which are relevant to discussions of, e.g., rotating nuclei and nuclear magnetic moment, are entirely fixed by the Lorentz symmetry \cite{Afanasjev2010a, Afanasjev2010b, Zhao2011PLB, Zhao2011PRL, Meng2013}, while those in the non-relativistic models may have some ambiguity \cite{Dobaczewski1995}.

Throughout this paper we employ a relativistic energy density functional based on the relativistic point-coupling model.
This is a zero-range interaction analogous to the non-relativistic Skyrme interaction.
The effective Lagrangian is given by
\begin{equation}
{\cal L}={\cal L}_{\rm free}+{\cal L}_{\rm em}+{\cal L}_{\rm int},
\end{equation}
where
\begin{equation}\label{eq:L_N}
{\cal L}_{\rm free}=\bar\psi(i\gamma^\mu\partial_\mu-m)\psi
\end{equation}
is the free nucleon part, with $\psi$ and $m$ being the nucleon field and nucleon mass, respectively.
The electromagnetic part ${\cal L}_{\rm em}$ is given by
\begin{equation}\label{eq:L_em}
{\cal L}_{\rm em}=-\frac{1}{4}F^{\mu\nu}F_{\mu\nu}-\bar\psi\frac{1-\tau_3}{2}\gamma^\mu A_\mu\psi,
\end{equation}
where $A^\mu$ is the electromagnetic field and $F^{\mu\nu}=\partial^\mu A^\nu-\partial^\nu A^\mu$ is the electromagnetic field strength tensor.
The nuclear interaction ${\cal L}_{\rm int}$ characterizes the relativistic point-coupling model.
It consists of the four-fermion couplings ${\cal L}_{\rm 4f}$, derivative terms ${\cal L}_{\rm der}$, and higher-order terms ${\cal L}_{\rm hot}$ \cite{Burvenich2002}:
\begin{equation}
{\cal L}_{\rm int}={\cal L}_{\rm 4f}+{\cal L}_{\rm der}+{\cal L}_{\rm hot}.
\end{equation}
Here, ${\cal L}_{\rm 4f}$ is the leading-order of zero-range approximation to the meson-exchange interaction \cite{Liang2012},
\begin{eqnarray}\label{eq:NN4f}
\begin{aligned}
{\cal L}_{\rm 4f}=
&
-\frac{1}{2}\alpha_S(\bar{\psi}\psi)(\bar{\psi}\psi)
-\frac{1}{2}\alpha_V(\bar{\psi}\gamma_{\mu}\psi)(\bar{\psi}\gamma^{\mu}\psi)\\
&
-\frac{1}{2}\alpha_{TS}(\bar{\psi}\vec\tau\psi)\cdot(\bar{\psi}\vec\tau\psi) \\
&
-\frac{1}{2}\alpha_{TV}(\bar{\psi}\gamma_{\mu}\vec\tau\psi)\cdot(\bar{\psi}\gamma^{\mu}\vec\tau\psi),
\end{aligned}
\end{eqnarray}
while ${\cal L}_{\rm der}$ is the next-to-leading-order terms, which simulate the finite range of the meson-exchange interactions,
\begin{eqnarray}\label{eq:NNder}
\begin{aligned}
{\cal L}_{\rm der}=
&
-\frac{1}{2}\delta_S(\partial_{\mu}\bar{\psi}\psi)(\partial^{\mu}\bar{\psi}\psi)
-\frac{1}{2}\delta_V(\partial_{\mu}\bar{\psi}\gamma_{\nu}\psi)
(\partial^{\mu}\bar{\psi}\gamma^{\nu}\psi)\\
&
-\frac{1}{2}\delta_{TS}(\partial_{\mu}\bar{\psi}\vec\tau\psi)\cdot(\partial^{\mu}\bar{\psi}\vec\tau\psi)\\
&
-\frac{1}{2}\delta_{TV}(\partial_{\mu}\bar{\psi}\gamma_{\nu}\vec\tau\psi)
\cdot(\partial^{\mu}\bar{\psi}\gamma^{\nu}\vec\tau\psi).
\end{aligned}
\end{eqnarray}
The ${\cal L}_{\rm hot}$ corresponds to the self-couplings of the scalar and vector mesons, which introduce a density dependence into the $N$-$N$ contact couplings,
\begin{equation}\label{eq:NNhot}
{\cal L}_{\rm hot}=
-\frac{1}{3}\beta_S(\bar{\psi}\psi)^3
-\frac{1}{4}\gamma_S(\bar{\psi}\psi)^4
-\frac{1}{4}\gamma_V\left[(\bar{\psi}\gamma_{\mu}\psi)
(\bar{\psi}\gamma^{\mu}\psi)\right]^2.
\end{equation}
Notice that the indices of the coupling constants, $S$, $V$, $TS$, and $TV$, denote the spin-isospin properties of the vertices, and they correspond to the exchanges of the $\sigma$, $\omega$, $\rho$, and $\delta$ mesons, respectively.

Under the assumption of time-reversal symmetry, the energy density functional is obtained from the Lagrangian with the Hartree and the no-sea approximations as
\begin{eqnarray}\label{eq:edf}
E&=&\int d^3r~
\Biggl[~\sum_{i=1}^A\psi^\dagger_i(-i\bvec\alpha\cdot\bvec\nabla+\beta m)\psi_i
+\frac{e}{2}\rho_V^pA^0\Biggr.\nonumber\\
&&~~~~~~\Biggl.+\frac{1}{2}\sum_{K}(\alpha_K\rho_K^2+\delta_K\rho_K\Delta\rho_K)\Biggr.\nonumber\\
&&~~~~~~\Biggl.+\frac{1}{3}\beta_S\rho_S^3+\frac{1}{4}\gamma_S\rho_S^4+\frac{1}{4}\gamma_V\rho_V^4~\Biggr],
\end{eqnarray}
where $K$ runs over the four spin-isospin channels, and the corresponding densities read
\begin{subequations}
\begin{eqnarray}
\rho_S(\bvec r)&=&\sum_{i=1}^{A}\bar{\psi}_i(\bvec r)\psi_i(\bvec r), \\
\rho_V(\bvec r)&=&\sum_{i=1}^{A}\psi_i^{\dagger}(\bvec r)\psi_i(\bvec r), \\
\rho_{TS}(\bvec r)&=&\sum_{i=1}^{A}\bar{\psi}_i(\bvec r)\tau_3\psi_i(\bvec r), \\
\rho_{TV}(\bvec r)&=&\sum_{i=1}^{A}\psi_i^{\dagger}(\bvec r)\tau_3\psi_i(\bvec r).
\end{eqnarray}
\end{subequations}
Here $\psi_i(\bvec r)$ is the wave function for the $i$-th nucleon.

The relativistic Hartree equation, or the relativistic Kohn-Sham equation, is obtained by taking a variation of the energy functional in Eq.~(\ref{eq:edf}) with respect to the single-particle wave functions as
\begin{eqnarray}
&&\left[-i\bvec\alpha\cdot\bvec\nabla+V_V+V_{TV}\tau_3+V_C
+\beta\left(m+V_S+V_{TS}\tau_3\right)\right]\psi_i \nonumber\\
&&=\epsilon_i\psi_i,
\end{eqnarray}
with
\begin{subequations}
\begin{eqnarray}
V_S&=&
\alpha_S\rho_S+\beta_S\rho_S^2+\gamma_S\rho_S^3+\delta_S\Delta\rho_S,\\
V_V&=&\alpha_V\rho_V+\gamma_V\rho_V^3+\delta_V\Delta\rho_V,\\
V_{TS}&=&\alpha_{TS}\rho_{TS}+\delta_{TS}\Delta\rho_{TS},\\
V_{TV}&=&\alpha_{TV}\rho_{TV}+\delta_{TV}\Delta\rho_{TV},\\
V_C&=&eA^0\frac{1-\tau_3}{2},\ (\Delta A^0=-e\rho_V^{(p)}),
\end{eqnarray}
\end{subequations}
where $\rho_V^{(p)}=\frac{1}{2}(\rho_V-\rho_{TV})$ is the proton density.
These equations are solved self-consistently to obtain the ground state of atomic nuclei.
Finally, one obtains the total binding energy as
\begin{eqnarray}
E_B&=&\sum_{i=1}^{A}\epsilon_i-E_{\rm CM}-Am \nonumber\\
&&
-\int d^3r\ \biggl(
\frac{1}{2}\sum_{K}\alpha_K\rho^2_K+\frac{1}{2}\sum_{K}\delta_K\rho_K\Delta\rho_K
\biggr.\\
&&
\biggl.
~~~~~~+\frac{2}{3}\beta_S\rho_S^3+\frac{3}{4}\gamma_S\rho_S^4+\frac{3}{4}\gamma_V\rho_V^4
+\frac{1}{2}eA^0\rho^{(p)}_V
\biggr),  \nonumber
\end{eqnarray}
where the center-of-mass energy $E_{\rm CM}$ is calculated by taking the expectation value of the kinetic energy for the center-of-mass motion of the whole nucleus with respect to the many-body ground-state wave function as \cite{Bender2000}
\begin{equation}\label{eq:cmenergy}
E_{\rm CM}=\frac{\langle \bvec P_{\rm CM}^2 \rangle}{2Am}.
\end{equation}

\section{Variational collapse in relativistic calculations}\label{sec:VC}

\subsection{Variational collapse}

Our aim in this paper is to carry out three-dimensional mesh calculations with the energy density functional given by Eq.~(\ref{eq:edf}).
To this end, we have to deal with the two challenges mentioned in Sec.~\ref{sec:intro}.
In this section, we consider the first problem, i.e., the variational collapse, and discuss the practical solutions for that.

The variational principle is a simple but powerful guiding principle to find approximate solutions 
in non-relativistic quantum mechanical problems.
According to the variational principle, one obtains a better solution by minimizing the energy as much as possible.
The imaginary time method, which has been successfully employed in 3D coordinate-space calculations for the non-relativistic systems, is entirely based on such a variational principle \cite{Davies1980}.
That is, the variational principle guarantees that the evolution in imaginary time,
\begin{equation}
\lim_{\tau\to\infty}\{e^{-h\tau}|\psi_k\rangle\},
\end{equation}
where $\{|\psi_k\rangle\}$ denotes a set of single-particle wave functions, decreases the total Hartree-Fock energy as a function of $\tau$, and eventually leading to a self-consistent solution.

In contrast, in the relativistic systems, the imaginary time evolution inevitably breaks down due to the presence of the Dirac sea states below the Fermi sea.
If one naively applies the imaginary time method, the iterative solution dives into the continuum in the Dirac sea, which has been numerically confirmed in Ref.~\cite{Zhang2010}.
This occurs since the imaginary time evolution seeks for the lowest single-particle states.
This is not what one wants, since only the lowest states in the Fermi sea are required in usual mean-field calculations with the no-sea approximation, whereas the Dirac sea states have to be explicitly excluded.

Such a breakdown of variational calculations, that is, 
the energy minimization, for relativistic systems is called variational collapse.
The variational collapse problem has long been known and discussed in the field of relativistic quantum chemistry \cite{Wallmeier1981, Wallmeier1983, Stanton1984, Hill1994, Falsaperla1997}, mainly in the case of basis-expansion calculations.
It has recently been discussed also
in the context of nuclear physics \cite{Zhang2009CPC, Zhang2009CPL, Zhang2010, Hagino2010}, in connection to a realization of CDFT calculations on 3D coordinate-space representation.

\subsection{Inverse Hamiltonian method}

\begin{figure}
\begin{center}
\includegraphics[width=8cm]{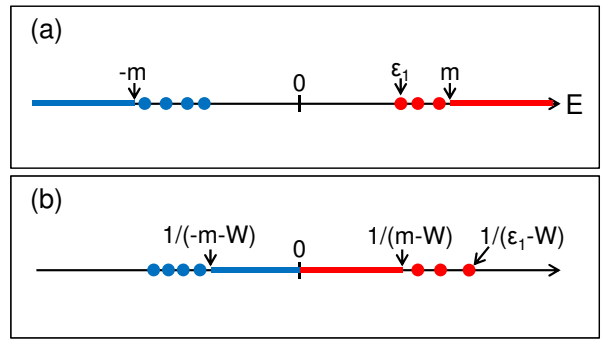}
\end{center}
\caption{(Color online) A schematic picture for a spectrum of (a) a relativistic single-particle Hamiltonian $h$ with mass $m$ and (b) its inverse $1/(h-W)$.
The positive- and negative-energy eigenvalues are denoted by red and blue colors, respectively.
The bound states are indicated by the solid circles while the continuum states are represented by the solid lines.}
\label{fig:rel-spectrum}
\end{figure}

In order to avoid the variational collapse in the relativistic calculations, we follow the idea of Hill 
and Krauthauser \cite{Hill1994}, which is based on the
variational principle for operator $1/(h-W)$, where $W$ is a real number not equal to any of the eigenvalues of $h$.
As shown in Fig.~\ref{fig:rel-spectrum}(a), the ordinary variational principle is not applicable to a relativistic Hamiltonian because it has a negative-energy spectrum down to negative infinity as well as a positive-energy spectrum up to positive infinity.
States between the two continua are discrete bound states.
On the other hand, in the spectrum of the inverse Hamiltonian, $1/(h-W)$, the two continua come to the middle of the spectrum while the bound states come to the two ends, see Fig.~\ref{fig:rel-spectrum}(b).
The positive (negative)-energy bound states come to the top (bottom) of the spectrum when the constant $W$ is set between the Fermi sea and the Dirac sea.

Let us label the energy eigenstates of the Hamiltonian with an integer $k$ according to the energy, so that the energy eigenvalues are denoted as
\begin{equation}
\ldots\leq\epsilon_{-2}\leq\epsilon_{-1}<\epsilon_1\leq\epsilon_{2}\leq\ldots,
\end{equation}
where $k>0$ ($k<0$) corresponds to the Fermi (Dirac) sea solutions, with $|\phi_k\rangle$ being the eigenstate associated to $\epsilon_k$.
For simplicity of notations, here we have treated the continuum states as discrete states.
Since the spectrum is bound both from the above and the below, we have the rigorous variational principle expressed as
\begin{equation}
\frac{1}{\epsilon_{-1}-W}\leq \frac{\langle\psi|(h-W)^{-1}|\psi\rangle}
{\langle\psi|\psi\rangle}\leq\frac{1}{\epsilon_1-W}.
\end{equation}
This implies that a maximization of the expectation value of $1/(h-W)$ leads to an approximate solution to the lowest positive-energy state, while a minimization leads to the state on the top of the Dirac sea.
In Ref.~\cite{Hill1994}, Hill and Krauthauser showed that this strategy indeed works for a variational calculation with trial functions defined with variational parameters.

Based on the variational principle of Hill and Krauthauser for the inverse of Hamiltonian, we have proposed in Ref.~\cite{Hagino2010} the inverse Hamiltonian method for an iterative solution of Dirac equations in the coordinate-space representation.
In this method, an initial state $|\psi^{(0)}\rangle$, which is not an eigenstate of the Hamiltonian, is evolved with an operator $e^{T/(h-W)}$ as
\begin{equation}\label{eq:invH0}
e^{T/(h-W)}|\psi^{(0)}\rangle=
\sum_k e^{T/(\epsilon_k-W)}|\phi_k\rangle\langle\phi_k|\psi^{(0)}\rangle.
\end{equation}
Note that $T$ has a dimension of energy and has nothing to do with time.
The initial state $|\psi^{(0)}\rangle$ can be always expanded formally with the true eigenstates of the Hamiltonian $h$.
In the evolution with $e^{T/(h-W)}$, all the negative-energy eigenstates contained in the initial state will damp away because the exponent is all negative.
On the other hand, all the positive-energy states will grow up exponentially since the exponent is all positive, among which the state closest to $W$ in the Fermi sea, $|\phi_1\rangle$, grows up most rapidly.
Therefore, in the limit of $T\to\infty$, the wave function
converges to the lowest state in the Fermi sea, i.e., \begin{equation}\label{eq:invH0T}
\lim_{T\to\infty}\sum_k e^{T/(\epsilon_k-W)}|\phi_k\rangle\langle\phi_k|\psi^{(0)}\rangle
\propto |\phi_{1}\rangle.
\end{equation}
Moreover, if one takes $T\to -\infty$, the wave function converges to the highest state in the Dirac sea.
Hereafter we only consider a positive $T$.
With this inverse Hamiltonian method, one can obtain exclusively the positive-energy solutions, which are usually of relevance to nuclear structure calculations.

One can obtain the higher-energy states by starting with a set of initial wave functions and orthonormalizing them during the evolution by the Gram-Schmidt method.
Alternatively, the iterative solution could also converge to $|\phi_2\rangle$ by setting the shift parameter $W$ to be between $\epsilon_1$ and $\epsilon_2$.
This scheme, however, may have a problem since the eigenvalues of $h$ are not known \textit{a priori}, and $1/(h-W)$ becomes singular when $W$ happens to be equal to one of them.

In practical calculations, one may cut $T$ into steps by $\Delta T$, and carry out the evolution iteratively.
That is, in every step, the exponential function $e^{\Delta T/(h-W)}$ is expanded to the first order of $\Delta T$.
The iterative wave function at $T=(n+1)\Delta T$ is then given with the wave function at the previous step as
\begin{equation}
|\psi^{(n+1)}\rangle\propto\left(1+\frac{\Delta T}{h-W}\right)|\psi^{(n)}\rangle.
\end{equation}
Since this evolution operator is not unitary, one has to normalize the wave function at each step.

One can show that the expectation value of the inverse of the Hamiltonian monotonically increases.
In order to demonstrate this, let us assume that the step size $\Delta T$ is sufficiently small so that its second order is negligible.
To the first order of $\Delta T$, one obtains
\begin{eqnarray}\label{eq:invH-evol}
&&\frac{\langle\psi^{(n+1)}|h^{-1}|\psi^{(n+1)}\rangle}{\langle\psi^{(n+1)}|\psi^{(n+1)}\rangle} \nonumber \\
&=&\frac{\langle\psi^{(n)}|\left(1+\frac{\Delta T}{h}\right)
\frac{1}{h}\left(1+\frac{\Delta T}{h}\right)|\psi^{(n)}\rangle}
{\langle\psi^{(n)}|\left(1+\frac{\Delta T}{h}\right)
\left(1+\frac{\Delta T}{h}\right)|\psi^{(n)}\rangle} \nonumber \\
&\simeq&
\frac{\langle\psi^{(n)}|h^{-1}|\psi^{(n)}\rangle+2\Delta T\langle\psi^{(n)}|h^{-2}|\psi^{(n)}\rangle}
{1+2\Delta T\langle\psi^{(n)}|h^{-1}|\psi^{(n)}\rangle} \nonumber \\
&\simeq&
\langle\psi^{(n)}|h^{-1}|\psi^{(n)}\rangle \nonumber \\
&&+2\Delta T\left[\langle\psi^{(n)}|h^{-2}|\psi^{(n)}\rangle
-\left(\langle\psi^{(n)}|h^{-1}|\psi^{(n)}\rangle\right)^2\right]. \nonumber \\
\end{eqnarray}
In the last line, the $\Delta T$ term is positive definite, since it is proportional to the dispersion of $h^{-1}$.
One therefore finds,
\begin{equation}
\frac{\langle\psi^{(n+1)}|h^{-1}|\psi^{(n+1)}\rangle}{\langle\psi^{(n+1)}|\psi^{(n+1)}\rangle}
\geq\langle\psi^{(n)}|h^{-1}|\psi^{(n)}\rangle.
\end{equation}
That is, the expectation value of $h^{-1}$ increases from the step $n$ to the step $n+1$.
This is a natural and reasonable consequence of the variational principle for the inverse of Hamiltonian.

Notice that, in contrast to $h^{-1}$, the behavior of the energy expectation value, $\langle h\rangle$, is not necessarily monotonic:
\begin{eqnarray}\label{eq:e-evol}
&&\frac{\langle\psi^{(n+1)}|h|\psi^{(n+1)}\rangle}{\langle\psi^{(n+1)}|\psi^{(n+1)}\rangle} \nonumber \\
&\simeq&\langle\psi^{(n)}|h|\psi^{(n)}\rangle \nonumber \\
&&+2\Delta T\left(1-\langle\psi^{(n)}|h|\psi^{(n)}\rangle\langle\psi^{(n)}|h^{-1}|\psi^{(n)}\rangle\right),
\end{eqnarray}
where the sign of the $\Delta T$ term depends on the property of $\psi^{(n)}$, and it can be either 
positive or negative.
Notice also that both in Eqs.~(\ref{eq:invH-evol}) and (\ref{eq:e-evol}),
the $\Delta T$ term in the last line converges to zero as the iterative wave function converges to an eigenstate of the Hamiltonian.

In Ref.~\cite{Hagino2010}, it has been shown that the inverse Hamiltonian method works well for eigenvalue problems of a radial Dirac equations with a given spherical potential.
This method is relatively straightforward to be applied to not only the Dirac equation but also other eigenvalue problems with unbound operators, which is one of its advantages over some other methods for variational collapse \cite{Wallmeier1983, Stanton1984, Zhang2009CPC, Zhang2009CPL, Zhang2010}.
In this context, we have successfully applied the inverse Hamiltonian method to non-relativistic Hartree-Fock-Bogoliubov equations with spherical potentials \cite{Tanimura2013}.

It is straightforward to extend the inverse Hamiltonian method to the self-consistent calculations.
To this end, we simply replace the imaginary time evolution for the non-relativistic mean-field calculations with the $T$-evolution given by Eq.~(\ref{eq:invH0}).
That is,
\begin{enumerate}
\item Prepare an initial set of single-particle wave functions $\{\psi_k^{(0)}\}$ with $k=1,2,...,A$.
\item Construct the density $\rho^{(0)}$ with $\{\psi_k^{(0)}\}$.
\item Construct the single-particle Hamiltonian $h^{(0)}$, which is a functional of the density $\rho^{(0)}$.
\item Generate a new set of the single-particle wave functions, $\{\tilde\psi_k\}$, by applying the $T$-evolution on each wave function:
    \begin{equation}
    |\tilde\psi^{(1)}_k\rangle = \exp[\Delta T/(h^{(0)}-W^{(0)})]|\psi_k^{(0)}\rangle,
    \end{equation}
\item Orthonormalize the set $\{\tilde\psi_k^{(1)}\}$ to obtain $\{\psi_k^{(1)}\}$. Go back to the step 2.
\end{enumerate}
The steps from 2 to 5 are iterated until the convergence is achieved.
One may change the energy shift $W^{(n)}$ at each iteration.

Here we closely follow the discussion in Ref.~\cite{Davies1980} in order to show that the iteration leads to
a self-consistent solution.
The evolution of the wave functions by a small step $\Delta T$ can be approximated by
\begin{equation}
|\tilde\psi_k^{(n+1)}\rangle=[1-\Delta T(h^{(n)}-W^{(n)})^{-1}]|\psi_k^{(n)}\rangle.
\end{equation}
With the Gram-Schmidt orthonormalization, one obtains the wave functions
\begin{eqnarray}
|\psi_k^{(n+1)}\rangle&=&\left\{1-\Delta T\left[\zeta_{kk}^{(n)}-(h^{(n)}-W^{(n)})^{-1}\right]\right\}
|\psi_k^{(n)}\rangle \nonumber \\
&&-2\Delta T\sum_{l<k}\zeta_{lk}^{(n)}|\psi_l^{(n)}\rangle,
\label{eq:spwf-GS}
\end{eqnarray}
where the matrix elements of the inverse Hamiltonian is denoted as $\langle\psi_l^{(n)}|(h^{(n)}-W^{(n)})^{-1}|\psi_k^{(n)}\rangle\equiv\zeta^{(n)}_{lk}$.
Since the change in the density matrix from the $n$-th step to the $(n+1)$-th step is given by
\begin{equation}
\rho^{(n+1)}-\rho^{(n)}=\sum_{k=1}^A\left(|\psi_k^{(n+1)}\rangle\langle\psi_k^{(n+1)}|
-|\psi_k^{(n)}\rangle\langle\psi_k^{(n)}|\right),
\end{equation}
one obtains
\begin{eqnarray}
\rho^{(n+1)}-\rho^{(n)}&=&
\Delta T\left[\rho^{(n)}\frac{1}{h^{(n)}-W^{(n)}}(1-\rho^{(n)})\right. \nonumber \\
&&\left. +(1-\rho^{(n)})\frac{1}{h^{(n)}-W^{(n)}}\rho^{(n)}\right],
\end{eqnarray}
which leads to
\begin{eqnarray}
&&{\rm Tr}\left[\frac{1}{h^{(n)}-W^{(n)}}(\rho^{(n+1)}-\rho^{(n)})\right] \nonumber\\
&=&
2\Delta T\cdot{\rm Tr}\left[\rho^{(n)}\frac{1}{h^{(n)}-W^{(n)}}(1-\rho^{(n)})
\frac{1}{h^{(n)}-W^{(n)}}\right].\nonumber\\
\end{eqnarray}
Defining $B^\dagger=\rho^{(n)}(h^{(n)}-W^{(n)})^{-1}(1-\rho^{(n)})$, one thus obtains
\begin{equation}
{\rm Tr}\left[\frac{1}{h^{(n)}-W}(\rho^{(n+1)}-\rho^{(n)})\right]
=2\Delta T
\cdot{\rm Tr}\left[B^\dagger B\right] > 0.
\end{equation}
Therefore, the quantity
\begin{equation}
{\rm Tr}\left[\frac{1}{h^{(n)}-W^{(n)}}\rho^{(n)}\right]=
\sum_{k=1}^A \zeta_{kk}^{(n)}
\end{equation}
always increases from iteration to iteration, which indicates that $\{|\psi_k^{(n+1)}\rangle\}$ are better solutions of $h^{(n)}$ than $\{|\psi_k^{(n)}\rangle\}$.
This quantity will eventually converge to its maximum value (for a fixed value of $W$), implying that $B^\dagger=0$, or
\begin{equation}
\left[\frac{1}{h-W},\rho\right]=0.
\end{equation}
Multiplying $h-W$ from right and left on both sides of the above equation, one finally obtains
\begin{equation}
[h,\rho]=0
\end{equation}
at the convergence.

\section{Fermion doubling problem}\label{sec:FD}

\subsection{Dispersion relation and Fermion doubling}

In the previous section, we have discussed one of the major problems, variational collapse, in the relativistic mean-field calculations on 3D lattice.
In this section we will discuss the other problem, i.e., the fermion doubling.
The fermion doubling problem has been well known in the field of lattice QCD, in which the first derivative in the action is replaced by a finite difference by discretizing the space-time within a box with the periodic boundary condition.
Here we shall show how the problem arises for a static one-particle Dirac equation.

For simplicity let us consider a Dirac equation for a free particle in 1-dimensional space,
\begin{equation}\label{eq:1dDirac_free}
(-i\alpha\partial_x+\beta m)\psi(x)=\epsilon\psi(x),
\end{equation}
where $\psi(x)$ is a two-component spinor and
\begin{equation}
\alpha=
\left(\begin{array}{cc}
0 & 1\\
1 & 0
\end{array}\right),\qquad
\beta=
\left(\begin{array}{cc}
1 & 0\\
0 & -1
\end{array}\right).
\end{equation}

Let us solve this equation by discretizing the coordinate $x$ with a mesh size $a$ within a box of size $L$.
To this end, we approximate the derivative in the kinetic term with the 3-point differential formula,
\begin{equation}
\partial_x\psi(x_i)=\frac{\psi(x_{i+1})-\psi(x_{i-1})}{2a},
\end{equation}
where $x_i\equiv a\cdot i$ is the $i$-th mesh point.
If either the periodic $\psi(x+L)=\psi(x)$ or the anti-periodic $\psi(x+L)=-\psi(x)$ boundary condition is imposed, the Dirac equation in the momentum space reads
\begin{equation}\label{eq:1d-dirac-k}
\left[\frac{1}{a}\alpha\sin(ka)+\beta m\right]\tilde\psi(k)=\epsilon\tilde\psi(k),
\end{equation}
where $\tilde\psi(k)$ is the Fourier transform of $\psi(x)$.
From Eq.~(\ref{eq:1d-dirac-k}) one obtains a dispersion relation
\begin{equation}\label{eq:disp-disc}
\epsilon^2=\frac{1}{a^2}\sin^2(ka)+m^2,
\end{equation}
which reduces to the ordinary dispersion relation of a relativistic particle
\begin{equation}\label{eq:disp-ex}
\epsilon^2=k^2+m^2
\end{equation}
in the continuum limit $a\to 0$.

\begin{figure}
\begin{center}
\includegraphics[angle=-90,width=8cm]{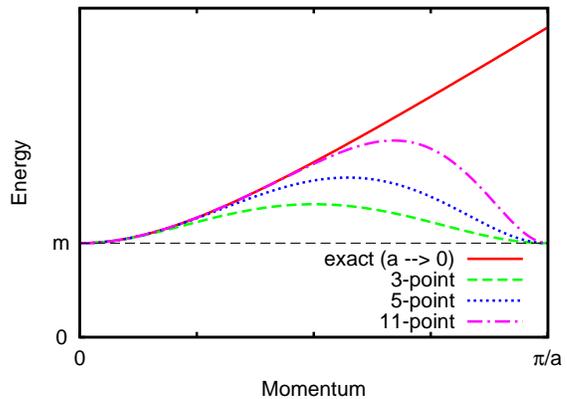}
\end{center}
\caption{(Color online) Dispersion relations for a free fermion in 1-dimension Dirac equation.
The solid curve shows the exact dispersion relation
given by Eq.~(\ref{eq:disp-ex}).
The dashed, dotted, and dash-dotted curves show the dispersion relation obtained with the 3-, 5-, and 11-point differential formulas for the kinetic term, respectively.}
\label{fig:fd}
\end{figure}

In Fig.~\ref{fig:fd}, we compare several approximate dispersion relations of a fermion on lattice with the exact one.
The solid curve is the exact dispersion relation in the continuum limit show in Eq.~(\ref{eq:disp-ex}), while the dashed curve shows the dispersion relation for a discretized Dirac equation given in Eq.~(\ref{eq:disp-disc}).
Moreover, the dotted and dash-dotted curves show the dispersion relations for a discretized lattice, obtained with the 5- and 11-point finite differential formulas, respectively.
In the continuum limit, energy increases monotonically as a function of momentum as shown by the solid curve.
In contrast, for the discretized fermions, a spurious minimum appears at the edge of the Brillouin zone, i.e., at
the cut-off momentum $k=\pi/a$ in the model space.
As seen in the figure, this minimum does not disappear even if more accurate point-difference formulas are used, although the lower-momentum behavior is improved.
If one solves the discretized Dirac equation, one thus
obtains not only physical solutions with low energy and low momentum, but also spurious solutions with {\it low energy and high momentum}.

In the 1D case, one spurious state appears for each physical state, while in the 3D cases there are seven spurious states for each physical solution.
In the 3D cases, spurious minima appear on each of three axes of 3D momentum space at the boundary of the Brillouin zone, and the seven spurious solutions correspond to
$(x_{\rm phys}, y_{\rm phys}, z_{\rm spur})$,
$(x_{\rm phys}, y_{\rm spur}, z_{\rm phys})$,
$(x_{\rm spur}, y_{\rm phys}, z_{\rm phys})$,
$(x_{\rm phys}, y_{\rm spur}, z_{\rm spur})$,
$(x_{\rm spur}, y_{\rm phys}, z_{\rm spur})$,
$(x_{\rm spur}, y_{\rm spur}, z_{\rm phys})$, and
$(x_{\rm spur}, y_{\rm spur}, z_{\rm spur})$,
where $x_{\rm phys}$ and $x_{\rm spur}$ denote the physical and spurious solutions in the $x$-direction, respectively, and similar for the $y$ and $z$ axes.

\subsection{Wilson fermion and its extension}\label{ssec:hoW}

\subsubsection{Wilson fermion}

In order to eliminate the spurious fermions, Wilson introduced a term proportional to $p^2$ into the action, which yields an additional contribution to the dispersion relation.
It can separate the energy of the spurious states on the high-momentum side from that of the physical states on the low-momentum side \cite{Wilson1977, Kogut1983}.
This method, referred to as the Wilson fermion, has been widely used in lattice QCD calculations.

We will demonstrate how it works for a static 1D Dirac equation.
With the 3-point formula for the kinetic energy, the action of an 1D Hamiltonian on a wave function $\psi(x)$ for a Wilson
fermion is defined as
\begin{eqnarray}
h_W\psi(x_i)&=&-i\alpha\frac{1}{2a}[\psi(x_{i+1})-\psi(x_{i-1})]+\beta m\psi(x_i) \nonumber\\
&-&\beta\frac{R}{a}[\psi(x_{i+1})-2\psi(x_i)+\psi(x_{i-1})].
\end{eqnarray}
The last term, the so-called Wilson term, has been added to the Hamiltonian.
Here $R$ is a dimensionless free parameter, and is referred to as a Wilson parameter.
The corresponding Hamiltonian in the continuum limit may be formally written as
\begin{equation}\label{eq:h_W1D}
h_W=-i\alpha\partial_x+\beta \left(m-aR\partial_x^2\right),
\end{equation}
which can be straightforwardly extended to the 3D space as
\begin{equation}
h_W=-i\bvec{\alpha}\cdot\bvec{\nabla}+\beta\left(m-aR\Delta\right).
\end{equation}
Notice that, in the continuum limit $a\to 0$, the Wilson term vanishes as a higher order, and the original form of Hamiltonian is recovered.

Since the Wilson term, $-aR\beta\Delta$, is proportional to $\bvec{p}^2$, this term lifts up the spurious minimum of the dispersion relation at the edge of the Brillouin zone.
Thus, the energy of the spurious states is all pushed upwards.
Even though the physical states are also affected by the Wilson term, the effect is much smaller than that for the spurious states, because the expectation values $\langle \bvec{p}^2\rangle$ are much smaller for the physical states.
Therefore, the doubling problem can be avoided if one takes the value of $R$ so that all the spurious states are pushed away from the energy region relevant to the calculation, that is,
the fermi energy in the case of mean-field calculations.
In Appendix, we discuss in detail the behavior of the energy spectrum of spurious single-particle states as a function of $R$.

\begin{figure}
\begin{center}
\includegraphics[angle=-90, width=8cm]{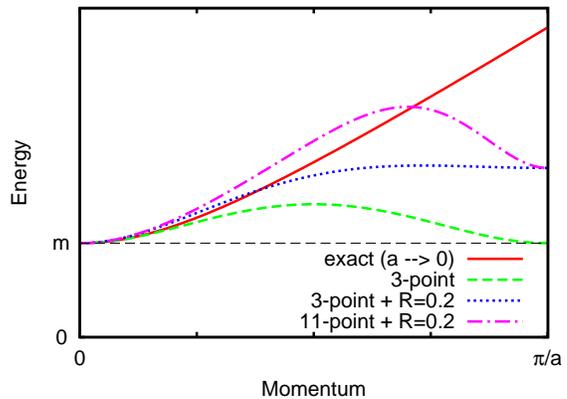}
\end{center}
\caption{(Color online) Comparison of the dispersion relations with and without the Wilson term (see Eq.~(\ref{eq:disp4})). 
The solid and dashed curves are the same as in Fig.~\ref{fig:fd}, while the dotted and dash-dotted curves show the dispersion relations with the Wilson term with $R=0.2$, obtained with the 3- and 11-point formulas for the kinetic energy term, respectively.}
\label{fig:wil-p2}
\end{figure}

For the 1D Hamiltonian given by Eq.~(\ref{eq:h_W1D}), the dispersion relation reads
\begin{equation}\label{eq:disp4}
\epsilon^2=\frac{1}{a^2}\sin^2(ka)+\left[m+\frac{2R}{a}(1-\cos(ka))\right]^2,
\end{equation}
when the 3-point formula is adopted for the kinetic term.
Notice that this also returns to the exact dispersion relation in Eq.~(\ref{eq:disp-ex}) in the continuum limit.
Figure~\ref{fig:wil-p2} shows a comparison of the dispersion relations with and without the Wilson term.
To this end, we use $R$=0.2.
For the comparison, the figure also shows the dispersion relation obtained with the 11-point formula (the dash-dotted line) as well as that in the continuum limit (the solid line).

As one can see, the spurious minimum at $k=\pi/a$ disappears due to the Wilson term.
At the same time, the dispersion relation deviates from the exact dispersion relation also on the low momentum side, although the effect vanishes at $k=0$.
That is, the Wilson term influences not only the spurious states but also the physical states, as may have been expected.

For instance, considering $a=0.8$~fm as a typical mesh size for practical 3D calculations, the fermi momentum of nucleons $k_F\simeq 1.36$~fm$^{-1}$ corresponds to $0.35 \pi/a$.
This implies that the dispersion relation is required to agree well with the exact one up to $k\simeq 0.35 \pi/a$.
Notice that the fermi energy of nucleons is $\epsilon_F=k_F^2\hbar^2/2m\simeq 38 {\rm MeV}\simeq 0.04m$, thus, it is sufficient for the mean-field-level calculations that the upward shift of the unphysical minimum by the Wilson term is $0.04m$ or larger.

In order to
see clearly
the effect of the Wilson term in the dispersion relation,
we have chosen the value $R=0.2$ rather
arbitrarily in Fig. \ref{fig:wil-p2}.
This value may be too large given that the shift of the minimum is comparable to $m$, much larger than a desired value $0.04m$.
We have, however, confirmed that the physical solutions are still affected by the Wilson term even we take a smaller value of $R$.

\subsubsection{High-order Wilson term}

It is desired to find a prescription which affects the physical solutions as small as possible and at the same time affects the spurious solutions as much as possible.
In the field of chemistry and solid-state physics, other prescriptions have also been proposed than the Wilson fermion
\cite{Busic1999, Fillion-Gourdeau2012, Hernandez2012}.
They are, however, not easy to apply when one employs an accurate differential formula, e.g., the 11-point formula, for the kinetic term.
Therefore, we here attempt to extend the Wilson fermion, which can be applied in a straightforward way even if the higher-point differential formula is adopted.

\begin{figure}
\begin{center}
\includegraphics[angle=-90, width=8cm]{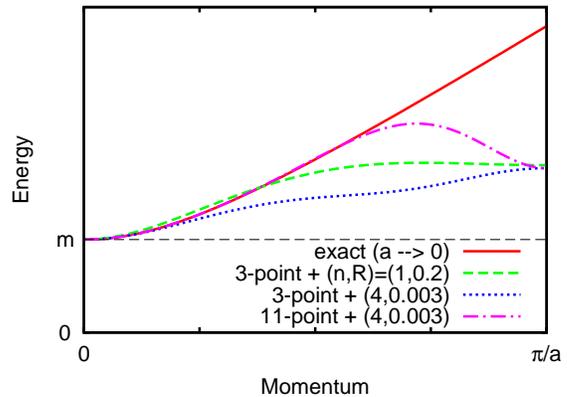}
\end{center}
\caption{(Color online) Comparison of the dispersion relations obtained with the high-order Wilson term (dotted and dash-dotted lines) to that with the original Wilson term (dashed line), as well as to the exact dispersion relation (solid line).
The parameters $(n,R)$ are defined in Eq.~(\ref{eq:imp-Wil}).
}
\label{fig:wil-p8}
\end{figure}

To this end, we increase the power of $p$ in the Wilson term as
\begin{equation}\label{eq:imp-Wil}
-aR\beta\sum_{i=1}^3\partial_i^2
\ \rightarrow\
(-)^{n}a^{2n-1}R\beta\sum_{i=1}^3\partial_i^{2n}.
\end{equation}
We show in Fig.~\ref{fig:wil-p8} the dispersion relations for the 1D Hamiltonian obtained with the prescription given by Eq.~(\ref{eq:imp-Wil}).
The dashed curve shows the result with the original Wilson fermion, corresponding to $(n,R)=(1,0.2)$ with the 3-point differential formula as in Fig.~\ref{fig:wil-p2}.
The dotted and dash-dotted curves show the improved dispersion relations obtained with $(n,R)=(4,0.003)$ together with the
3-point and 11-point differential formulas, respectively.
Evidently, the improved $(n>1)$ Wilson term leads to both a good agreement with the dispersion relation in the continuum limit in the low-$k$ region and an efficient separation of the spurious states from the physical states in the high-$k$ region.

\section{Self-consistent calculations}\label{sec:SC}

In the previous sections, we have introduced the two methods to overcome the variational collapse and fermion doubling problems.
We now combine them together and carry out the self-consistent relativistic mean-field calculations in the 3D coordinate-space representation.
To this end, we first perform a self-consistent calculation with the Wilson term using the inverse Hamiltonian method with an operator $e^{T/(h_W-W)}$, i.e.,
\begin{equation}
\lim_{T\to\infty}\left\{\exp[T/(h_W-W)]|\psi^{(0)}_k\rangle\right\}.
\end{equation}
After the convergence is achieved, we correct the energy by the first-order perturbation theory, that is,
the expectation value of the Wilson term is subtracted from the single-particle energies as well as from the total energy,
\begin{equation}
\epsilon_k =\tilde{\epsilon}_k-\epsilon_W=\tilde{\epsilon}_k
-\left\langle R\beta(-)^na^{2n-1}\sum_{i=1}^3\partial_i^{2n}\right\rangle,
\end{equation}
and
\begin{eqnarray}\label{eq:e-ew}
E&=&\tilde{E}-E_W \nonumber \\
&=&\tilde{E}-\left\langle R(-)^na^{2n-1}\sum_{k=1}^A\beta_k\sum_{i=1}^3
\partial_{ki}^{2n}\right\rangle,
\label{eq:Etotcorrect}
\end{eqnarray}
where $\tilde{\epsilon}_k$ and $\tilde{E}$ are the converged single-particle and total energies with the Wilson term, respectively.
In Eq.~(\ref{eq:Etotcorrect}), the subscript $k$ in $\beta_k$ and $\partial_{ki}$ means that they act on the $k$-th
single-particle state.
For simplicity, we do not correct the single-particle wave functions.

In the following, we present the results of benchmark calculations in order to show the validity of this strategy.
To this end, we perform calculations in a cubic numerical box with $L=a\times(N-1)$ long, where $a$ is the mesh size and $N$ is the number of mesh points taken along each direction.
We shall vary the box size $L$ and mesh size $a$, and examine the convergence of the results.
In the inverse Hamiltonian method, we need to invert a large sparse matrix of the single-particle Hamiltonians in the coordinate-space representation.
For this purpose, we employ an iterative solver, that is,
the conjugate residual method \cite{Saad2003}, for large sparse linear systems.
The Wilson parameter is fixed with $(n, R)=(5, 0.00015)$ for all the calculations shown in this section.
The first derivative of the kinetic term is approximated by the 11-point formula.
We take the box boundary condition, although the discussion of the dispersion relation was made
in Sec.~\ref{sec:FD} with the periodic boundary condition.
We have confirmed that the two boundary conditions yield almost the same results if the box size is large enough compared to the size of the system.

We first consider the $^{16}$O nucleus without the Coulomb interaction as an example.
We shall apply our method also to deformed nuclei.
We employ the energy functional of PC-F1 interaction given in Ref.~\cite{Burvenich2002}.
The self-consistency of the iterative solutions is judged by a condition that the dispersions of energy of all the occupied single-particle states are smaller than $10^{-8}$~MeV$^2$.

Since we discretize the full 3D space without assuming any spatial symmetry, we must constrain the center of mass of the whole nucleus at the origin and also the principal axes aligned to the coordinate axes:
\begin{equation}\label{eq:cnst-com}
\langle x\rangle\equiv 
\int d^3r\ x\rho_V(\bvec r)
=0,\quad \langle y\rangle=0,\quad {\rm and}\
\langle z\rangle=0,
\end{equation}
as well as
\begin{equation}\label{eq:cnst-pa}
\langle xy\rangle=0,\quad \langle yz\rangle=0,\quad {\rm and}\ \langle zx\rangle=0.
\end{equation}
When we draw potential energy surfaces, we in addition impose the constraints on the deformation parameters.
Here we define the nuclear multipole deformation parameters as
\begin{equation}\label{eq:defparam}
\alpha_{\ell m}=\frac{4\pi}{3AR^\ell}
\int d^3r\ r^\ell X_{\ell m}(\hat{\bvec r})\rho_V(\bvec r),\quad
-\ell\leq m \leq\ell,
\end{equation}
with $R=1.2\times A^{1/3}$~fm, where $X_{\ell m}$ is a real basis of the spherical harmonics,
\begin{equation}
X_{\ell m}=
\begin{cases}
Y_{\ell 0} & (m=0)\\
\frac{1}{\sqrt{2}}(Y_{\ell,-m}+Y_{\ell,-m}^*)& (m>0) \\
\frac{1}{\sqrt{2}i}(Y_{\ell,-m}-Y_{\ell,-m}^*)& (m<0)
\end{cases}.
\end{equation}
The ordinary quadrupole deformation parameters, $\beta$ and $\gamma$, are related to $\alpha_{2m}$ as $\alpha_{20}=\beta\cos\gamma$ and $\sqrt{2}\alpha_{22}=\beta\sin\gamma$.
We employ the augmented Lagrangian method \cite{Staszczak2010} for imposing these constraints.

\subsection{Box size dependence}

\begin{figure}
\begin{center}
\includegraphics[angle=-90, width=8cm]{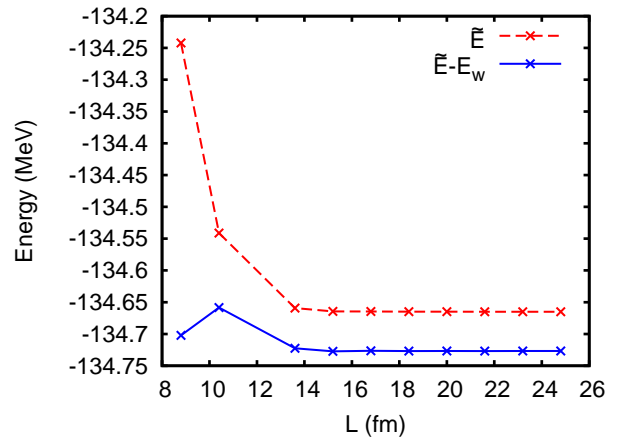}
\end{center}
\caption{(Color online) Convergence feature of the total binding energy of $^{16}$O with respect to the box size $L$.
The mesh size is fixed to be $a=0.8$~fm.
The Coulomb interaction and the center-of-mass correction are not included.
The dashed line shows the energies $\tilde{E}$ before the correction for the Wilson term, while the solid line shows the results after the correction, $E=\tilde{E}-E_W$.}
\label{fig:etot-L}
\end{figure}

We first discuss the dependence on the box size $L$.
Figure~\ref{fig:etot-L} shows the total energy of the $^{16}$O nucleus for various values of $L$, without the Coulomb interaction and the center-of-mass correction, while the mesh size is fixed to be $a=0.8$~fm.
The dashed line shows the energy $\tilde{E}$ before the correction of the Wilson term, while the solid is obtained after the correction, $\tilde{E}-E_W$ in Eq.~(\ref{eq:e-ew}).
Thus the difference between the two lines is nothing but the the expectation value of the Wilson term.
One can see that $L\simeq 15$~fm leads to a well converged result for $^{16}$O, while $E_W$ remains less than $0.1$~MeV.

\begin{figure}
\begin{center}
\includegraphics[angle=-90, width=8cm]{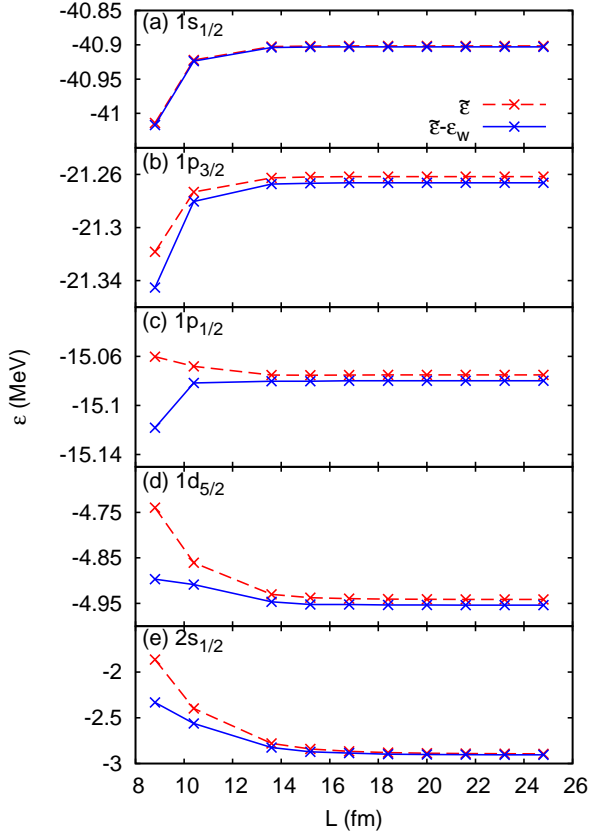}
\end{center}
\caption{(Color online) Same as Fig.~\ref{fig:etot-L}, but for the single-particle energies.}
\label{fig:esp-L}
\end{figure}

Figure~\ref{fig:esp-L} shows five single-particle energies as functions of $L$.
Even though $^{16}$O is a spherical nucleus, each multiplet of single-particle levels
may split in energy in the 3D mesh calculation with the Wilson term for several reasons
(see Sec.~\ref{ssec:oh} below).
Here we have averaged the energies within the same multiplet.
For the occupied states, i.e., the $1s_{1/2}$, $1p_{3/2}$, and $1p_{1/2}$ states, one sees that $L\simeq 15$~fm is sufficiently large to have a convergence, which is consistent with the results for the total energy shown in Fig.~\ref{fig:etot-L}.
On the other hand, a larger box size is required for the less bound single-particle states, e.g., $1d_{5/2}$ and $2s_{1/2}$, because of a long tail in these wave functions.

\subsection{Mesh size dependence}

We next discuss the convergence with respect to the mesh size $a$.
To this end, we fix the box size to be $L=a\times(N-1)\simeq 25$~fm.
For each value of $a$, we tune the number of mesh points $N$ so that $L$ is kept to be close to $25$~fm.
This value of $L$ yields well converged results for all the bound single-particle energies as shown in Fig.~\ref{fig:esp-L}, although $L\simeq 15$~fm may be sufficient for $^{16}$O.

\begin{figure}
\begin{center}
\includegraphics[angle=-90, width=8cm]{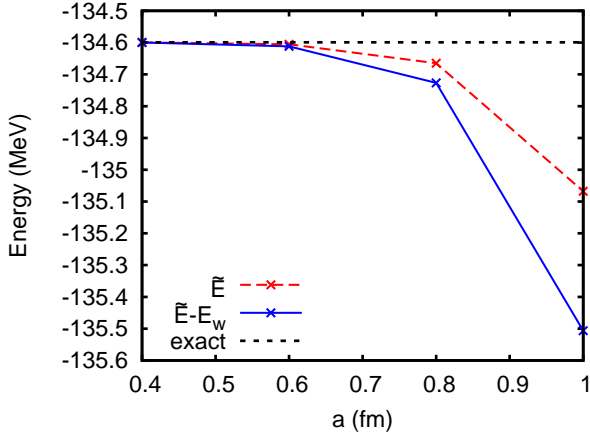}
\end{center}
\caption{(Color online) Convergence feature of the total binding energy of $^{16}$O with respect to the mesh size $a$.
The box size is kept to be $L\simeq 25$~fm for each value of $a$.
The Coulomb interaction and the center-of-mass correction are not included.
The meaning of the solid and dashed lines is
the same as in Fig.~\ref{fig:etot-L}, while the dotted line indicates the ``exact'' value of total energy obtained by a spherical code with the Runge-Kutta method.}
\label{fig:etot-a}
\end{figure}

Figure~\ref{fig:etot-a} shows the total binding energy of $^{16}$O obtained as a function of $a$.
For a comparison, the figure also shows by the dotted line the ``exact'' value obtained with a spherical code, in which the radial Dirac equation is solved by the Runge-Kutta method.
One can clearly see that the total energy converges to the exact value as the mesh size decreases.
Furthermore, the correction energy for the Wilson term becomes practically negligible for $a=0.6$~fm or smaller.
This is because the Wilson term is proportional to $a^{10}$ in this calculation and vanishes in the continuum limit.

Notice that the total energy increases as the mesh size $a$ becomes smaller.
This is opposite to what one would have expected from the variational principle for the non-relativistic systems.
This is, however, not surprising since there is no variational principle in the usual sense for the relativistic systems.
A similar phenomenon has been observed also in a relativistic mean-field calculation with basis expansion \cite{Lu2014MDC}.
That is, the total energy of a nucleus increases as the number of basis for the lower component of the Dirac spinor increases.
This is due to the variational collapse \cite{Stanton1984}, 
that is, when a basis for the lower component is increased by one, one state will be added to the negative-energy single-particle spectrum and it will push upwards all the states in the Fermi sea.
On the other hand, if a basis for the upper component is increased, a new state will appear on the top of the positive energy spectrum and it will push downwards all the other states.

\begin{figure}
\begin{center}
\includegraphics[angle=-90, width=8cm]{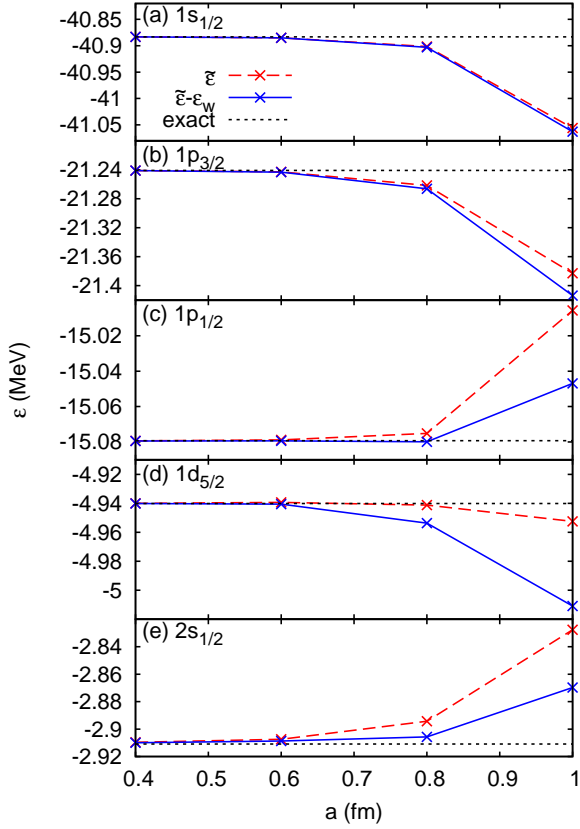}
\end{center}
\caption{(Color online) Same as Fig.~\ref{fig:etot-a}, but for the single-particle energies.}
\label{fig:esp-a}
\end{figure}

Figure \ref{fig:esp-a} shows the single-particle energies as functions of $a$.
As in Fig.~\ref{fig:esp-L}, we have taken the average for each multiplet.
It can be clearly seen that not only the total energy but also the single-particle energies approach closely to the exact results as the mesh size $a$ decreases.

\subsection{Accuracy of wave functions}

We next demonstrate that the wave functions obtained with the present 3D code are as accurate as the total and the single-particle energies.
To this end, we compare the total density of $^{16}$O obtained with the 3D code to that with the spherical code.
For the 3D mesh calculations, we use the box size of $L=25$~fm, the mesh size $a=0.6$~fm, and the 11-point formula for the kinetic term.
This is the same set up as that for the results shown in Figs.~\ref{fig:etot-a} and \ref{fig:esp-a}.

\begin{figure}
\begin{center}
\includegraphics[angle=-90, width=8cm]{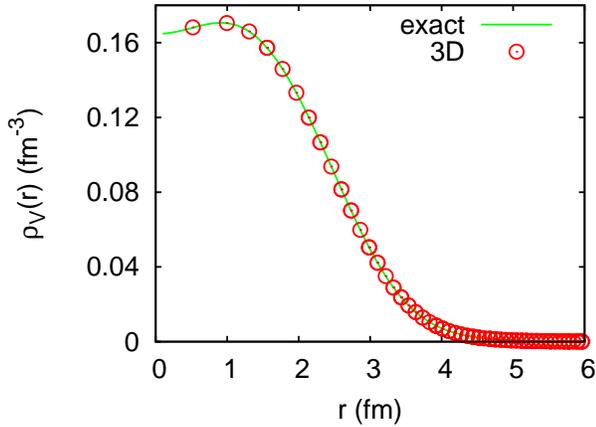}
\end{center}
\caption{(Color online)  The total density of $^{16}$O obtained with the 3D code (the open circles) and the spherical code (the solid line).
The Coulomb interaction is switched off.
The 3D calculation is carried out with the box size of $L=25$~fm, the mesh size $a=0.6$~fm, and the 11-point formula for the kinetic term.}
\label{fig:rho-r}
\end{figure}

\begin{figure}
\begin{center}
\includegraphics[angle=-90, width=8cm]{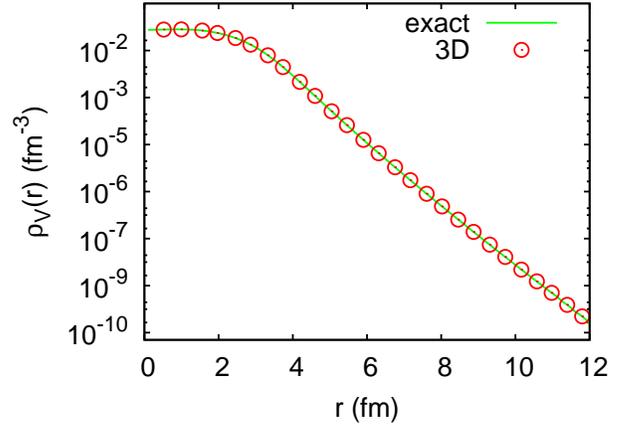}
\end{center}
\caption{(Color online)
Same as Fig.~\ref{fig:rho-r}, but in the logarithmic scale.}
\label{fig:lrho-r}
\end{figure}

Figure \ref{fig:rho-r} shows the total density distributions obtained with the two numerical codes, without the Coulomb interaction.
The data points in the 3D result is dense for large values of $r$, because there are many mesh points which correspond to similar values of $r$.
From the figure, one sees that the 3D code yields an almost identical result to the spherical code.
In Fig.~\ref{fig:lrho-r}, we make a similar comparison but in the logarithmic scale.
Notice that the range of $r$ is wider than in Fig.~\ref{fig:rho-r}.
One can see that the 3D code yields a quite accurate result up to very large $r$, including the logarithmic tail.
This is a particular nice feature of the real-space representation, whereas a basis representation may not be efficient in describing the asymptotic tail of the wave functions.

\subsection{Artificial violation of the rotational symmetry}\label{ssec:oh}

In the 3D mesh calculations with the Wilson term, the rotational symmetry is broken for several reasons.
Firstly, the high-order Wilson term introduced in Sec.~\ref{ssec:hoW} explicitly violates the rotational symmetry for $n\geq2$.
The original $SO(3)$ symmetry is broken down to the octahedral symmetry $O_h$.
Secondly, the lattice discretization, finite-difference approximation,
and the finite volume effect are also the sources of artificial violation of the rotational symmetry.
That is, both a cubic lattice and a cubic box reduce the symmetry to $O_h$, similarly to the high-order Wilson term.
The effects of the high-order Wilson term, the discretization, and the finite-difference
are expected to vanish in the continuum limit $a\to 0$, while the effect of finite volume vanishes with an infinitely large box $L\to\infty$.
In the actual calculations, one needs to take a sufficiently small $a$ and a large $L$ so that these artificial symmetry breaking can be neglected.

\subsubsection{Hexadecapole deformation}

\begin{table}
\caption{Hexadecapole deformation parameters $\alpha_{40}$ and $\alpha_{44}$ for $^{16}$O obtained with the self-consistent solutions of the 3D mesh calculations.
The results are shown for several different box sizes $L$, with fixed mesh size $a=0.8$~fm.
For all these cases, the corresponding quadrupole, octupole, and the other components of hexadecapole deformations are smaller
than 5$\times$10$^{-5}$.}
\begin{center}
\begin{tabular}{ccc}
\hline\hline
   $L$ (fm) & $\alpha_{40}$ & $\alpha_{44}$ \\
\hline
    $\,\,\, 8.8$ & $-0.0226$ & $-0.0191$ \\
          $10.4$ & $-0.0098$ & $-0.0083$ \\
          $13.6$ & $-0.0018$ & $-0.0015$ \\
          $16.8$ & $-0.0005$ & $-0.0004$ \\
          $20.0$ & $-0.0003$ & $-0.0003$ \\
          $23.2$ & $-0.0003$ & $-0.0003$ \\
\hline\hline
\end{tabular}
\end{center}
\label{tab:hex-L}
\end{table}

Let us discuss the violation of the rotational symmetry in the actual calculations, by taking again $^{16}$O without the Coulomb interaction as an example.
Since the ground state of this nucleus is spherical, all the deformation parameters are expected to vanish in principle.
In the actual calculations, however, the octahedral symmetry $O_h$, that is inherent in the 3D mesh calculation,
may induce spurious hexadecapole deformation.
Indeed, we find that the $\alpha_{40}$ and $\alpha_{44}$ in Eq.~(\ref{eq:defparam}) are small but finite in the self-consistent solutions.
Table~\ref{tab:hex-L} shows the values of $\alpha_{40}$ and $\alpha_{44}$ for $^{16}$O obtained with several different box sizes $L$, while the mesh size is kept to be $a=0.8$ fm.
Meanwhile, the other hexadecapole deformation parameters, as well as the quadrupole and octupole deformation parameters, are all less than $5\times 10^{-5}$, which are not shown in the Table.
The deformation parameters decrease as $L$ increases, and eventually converge to a small value, which is practically negligible.
The finite values which still remain even for large $L$ are due to the effects of discretization,
finite-difference, and the high-order Wilson term.

\begin{table}
\caption{Same as Table \ref{tab:hex-L}, but for different mesh size $a$.
The box size is kept to be $L\simeq 25$~fm.}
\begin{center}
\begin{tabular}{ccc}
\hline\hline
   $a$ (fm) & $|\alpha_{40}|$ & $|\alpha_{44}|$ \\
\hline
$1.0$ & $0.0025$ & $0.0021$ \\
$0.8$ & $0.0003$ & $0.0003$ \\
$0.6$ & $0.0001$ & $0.0001$ \\
$0.4$ & $< 5\times10^{-5}$ & $<5\times10^{-5}$ \\
\hline\hline
\end{tabular}
\end{center}
\label{tab:hex-a}
\end{table}

In Table~\ref{tab:hex-a}, we show the hexadecapole deformation parameters as a function of the mesh size $a$, while the box size is kept to be large enough, $L\simeq 25$ fm.
We see that the deformation becomes smaller as $a$ decreases, i.e., as it approaches the continuum limit, where the effects of discretization, finite-difference, and the high-order Wilson term vanish.
The hexadecapole deformations due to the artificial symmetry breaking are already negligibly small around $a=0.8$~fm, which is a typical mesh size in the 3D calculations.

\subsubsection{Splitting of single-particle levels}

Other quantities which are affected by the symmetry
violation are the single-particle energies.
With the group theory, it can be demonstrated that, for $j\geq 5/2$, the $(2j+1)$-fold degeneracy for the magnetic quantum numbers of the single-particle levels splits into several levels due to the reduction of symmetry from $SO(3)$ to $O_h$ \cite{Tanimura2014PhD}.
Similar level splitting of $SO(3)$ multiplet on 3D lattice was also discussed recently in nuclear physics context by Lu {\it et al.} in Ref.~\cite{Lu2014-3D}.
We thus investigate here the splitting of the $1d_{5/2}$ level in $^{16}$O as a function of the mesh size $a$, while the box size is kept to be $L\simeq25$~fm.

\begin{table}
\caption{Single-particle energies $\epsilon$ of $^{16}$O for the levels corresponding to $1d_{5/2}$ and $1p_{3/2}$ in the $SO(3)$ limit for several values of the mesh size $a$.
The values of $\epsilon$ and $a$ are given in units of MeV and fm, respectively.
The box size is set to be $L\simeq25$ fm.}
\begin{center}
\begin{tabular}{ccrrrr}
\hline\hline
$(n\ell j)$ & $|j_z|$ & $\epsilon(a=1.0)$ & $\epsilon(a=0.8)$
& $\epsilon(a=0.6)$ & $\epsilon(a=0.4)$ \\
\hline
$1d_{5/2}$ & $1/2$ & $-4.99669$ & $-4.95339$ & $-4.94013$ & $-4.94011$ \\
& $3/2$ & $-5.03357$ & $-4.95415$ & $-4.94134$ & $-4.94016$ \\
           & $5/2$ & $-5.00267$ & $-4.95354$ & $-4.94037$ & $-4.94012$ \\
\hline
$1p_{3/2}$ & $1/2$ & $-21.41360$ & $-21.26630$ & $-21.24336$ & $-21.24098$ \\
& $3/2$ & $-21.41360$ & $-21.26630$ & $-21.24336$ &  $-21.24098$  \\
\hline\hline
\end{tabular}
\end{center}
\label{tb:splevel}
\end{table}

In Table~\ref{tb:splevel}, we show the calculated single-particle energies of the levels corresponding to the $1d_{5/2}$
and $1p_{3/2}$ states in the $SO(3)$ limit for several values of the mesh size $a$.
As expected from the group theory, the $1d_{5/2}$ level splits in energy, whereas the $1p_{3/2}$ level does not within the digits shown in the table.
Similarly to the case of the hexadecapole deformation, the level splitting becomes smaller as the mesh size $a$ decreases, approaching to the $SO(3)$ limit.

We note that, with the symmetry violation, the $z$-component of the single-particle angular momentum, $j_z$, is no longer a good quantum number, and its actual expectation values are only approximately equal to the half-integers.
Since we prepare the initial single-particle wave functions on the lattice as spherical spinors, which are the eigenstates of $j_z$, the $j_z$ values are approximately conserved during the iteration, in which the axial symmetry is violated only slightly.
We also notice that, whereas the exact eigenstates of the mean-field potential with $O_h$ symmetry are the irreducible representation of the symmetry group, the resultant single-particle states in our present calculations are not the pure irreducible representations of $O_h$.
In this sense, the single-particle states with $j\geq 5/2$ in the present calculations cannot completely be the eigenstates of the mean-field Hamiltonian.
However, in practice, the convergence and the accuracy of the single-particle energies are satisfactory, as we have seen in this section.

\subsection{Deformed nuclei}

\begin{figure}
\includegraphics[angle=-90, width=8cm]{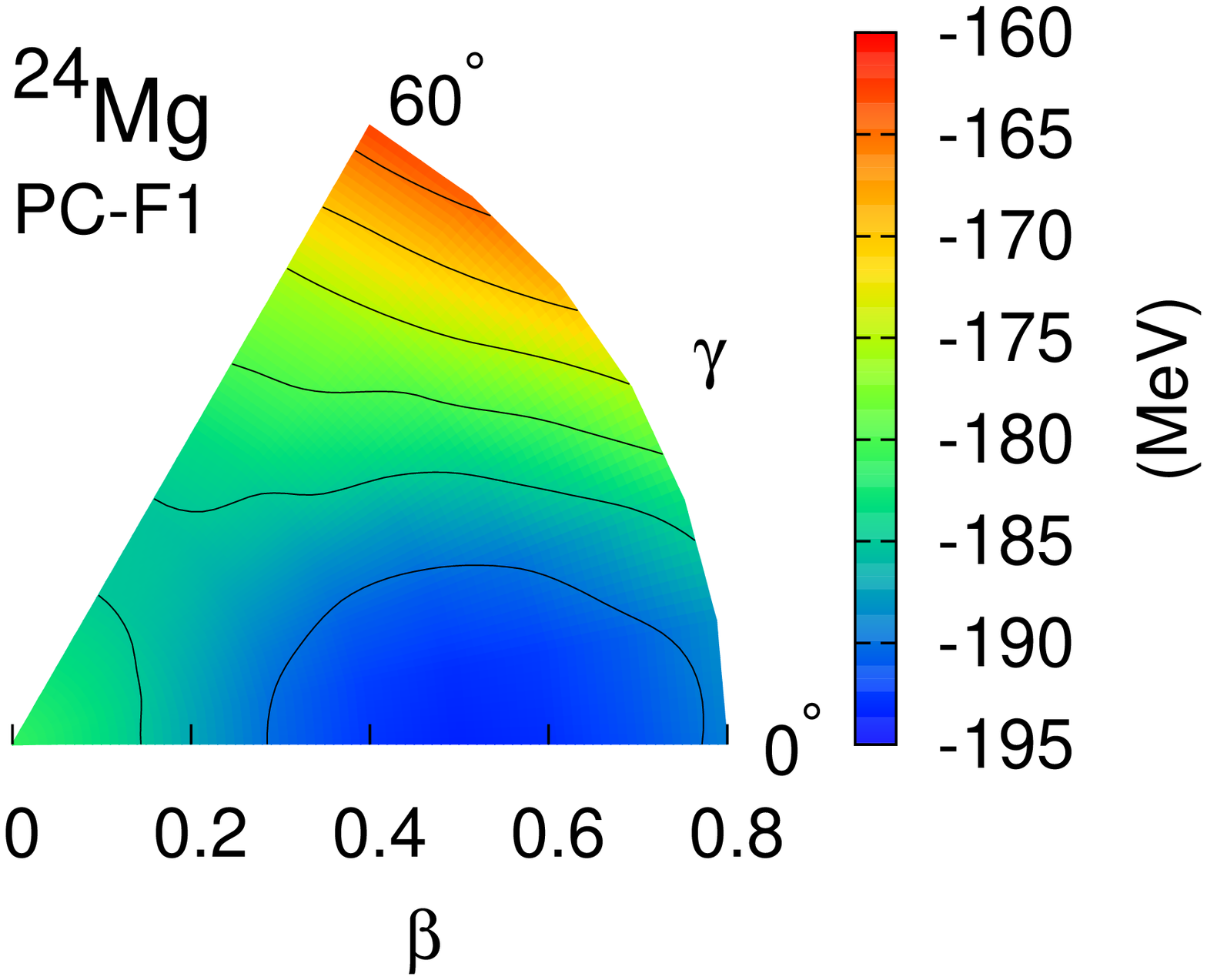}\\
\includegraphics[angle=-90, width=8cm]{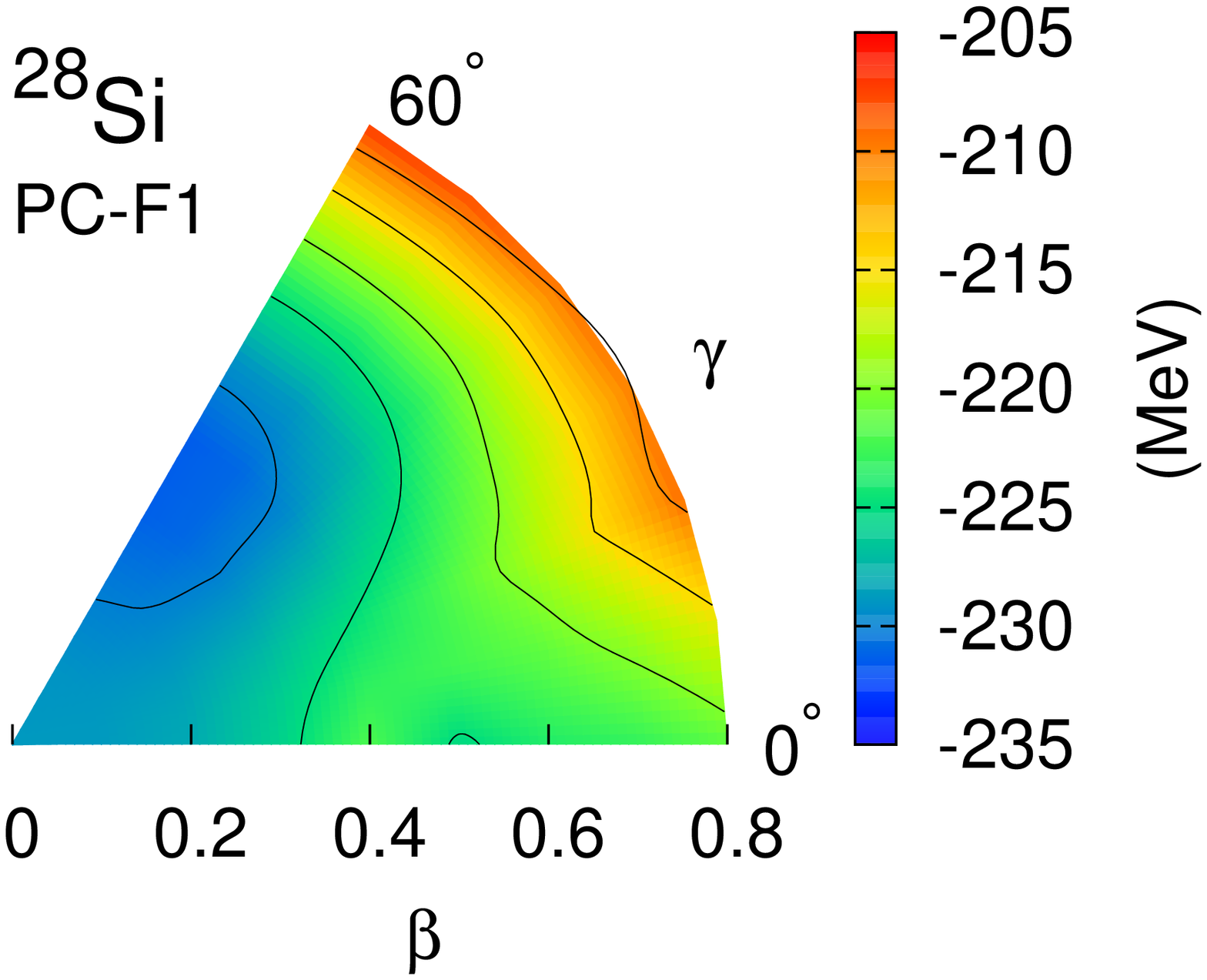}
\caption{(Color online) Potential energy surfaces of $^{24}$Mg (the upper panel) and $^{28}$Si (the lower panel) on the
$(\beta,\gamma)$ plane obtained with the 3D mesh CDFT calculations.}
\label{fig:PESbg}
\end{figure}

In the previous subsections, we have shown
that our method yields accurate solutions in the 3D coordinate-space representation for the spherical $^{16}$O nucleus, without the problems of variational collapse and fermion doubling.
Let us now apply the method to deformed nuclei, $^{24}$Mg and $^{28}$Si.
To this end, we take $20\times20\times20$ mesh points with $a=0.8$~fm.
Figure~\ref{fig:PESbg} shows the potential energy surface of $^{24}$Mg and $^{28}$Si on the $(\beta,\gamma)$ deformation plane obtained with the present 3D CDFT code.
One can see that a prolate deformation for $^{24}$Mg as well as an oblate deformation for $^{28}$Si are successfully reproduced in these calculations.
We emphasize that this is the first CDFT calculations on 3D lattice with constraints on deformation parameters.

\section{Summary and perspectives}\label{sec:sum}

Because of the variational collapse and the fermion doubling, a 3D coordinate-space calculation with covariant density functionals has been impossible for a long time.
In order to realize such calculations for the first time, we have proposed a novel and practical method to solve Dirac equations in the 3D coordinate space.
To this end, we have introduced the two different prescriptions and combined them to overcome the variational collapse and the fermion doubling.
For the variational collapse, we have employed a method based on the variational principle for the inverse of a single-particle Hamiltonian, while for the fermion doubling, we have extended the method of Wilson fermion, which has been widely employed in lattice QCD calculations.

Using $^{16}$O as an example, we have confirmed that our strategy provides accurate solutions for self-consistent mean-field calculations without the influence of the negative-energy spectrum and the spurious solutions of a discretized Dirac equation.
We have shown with $^{24}$Mg and $^{28}$Si that this method is also applicable to deformed solutions in the $(\beta,\gamma)$ deformation plane.
Such calculations are important for discussions of deformation properties as well as fission properties of heavy nuclei.
They can also be used as inputs for the generator coordinate method and five-dimensional collective Hamiltonian calculations in order to investigate spectroscopic properties of nuclei by taking into account the quantum fluctuation of shape degrees of freedom
\cite{Ring1980, Niksic2011,Yao14}.

There will be many possible applications of our new code.
In particular, it will enable one to
i) study any complicated structure of nuclei with a single numerical code,
ii) compare directly the results of the relativistic models to those of 3D mesh calculations with the non-relativistic models,
and iii) provide reliable theoretical predictions with the relativistic models for unknown nuclei allowing symmetry-breaking
solutions.
We emphasize that our new relativistic 3D code allows one to study arbitrary shape of nuclei such as exotic deformations, halo structure, complicated shape relevant to nuclear fission, and even a cluster structure, without any restriction on the spatial symmetry and without significantly increasing the numerical cost.
It also allows a straightforward extension of the finite-amplitude method \cite{Nakatsukasa2007, Nakatsukasa2012} within the relativistic framework \cite{Liang2013, Liang2014} for a study
of nuclear excitations in deformed nuclei.
We therefore believe that the method proposed in this work makes an important step to drastically extend the flexibility of the CDFT calculations in near future.

\begin{acknowledgments}
We thank M. Matsuo, J. Meng, P. Ring, H. Sagawa, S. Sasaki, J.M. Yao, Y. Zhang, and S.-G. Zhou for useful discussions.
This work was partly supported by the Grant-in-Aid for JSPS Fellows under the program numbers 24$\cdot$3429 and 24$\cdot$02201,
the Japanese Ministry of Education, Culture, Sports, Science and Technology by Grant-in-Aid for Scientific Research under the program number (C) 26400263,
and the RIKEN Foreign Postdoctoral Researcher Program.
The work of Y. T. was also supported by the Japan Society for Promotion of Science for Young Scientists.
\end{acknowledgments}

\appendix

\section{Fermion doubling in 1-dimensional space}

In this appendix, we will show an example of the fermion doubling problem and examine our prescription given in Sec.~\ref{sec:FD} in detail.

For this purpose, we consider a Dirac equation with the scalar $S(x)$ and vector $V(x)$ potentials in a 1-dimensional space,
\begin{equation}\label{eq:1ddirac_WS}
[-i\alpha\partial_x+V(x)+\beta(m+S(x))]\psi(x)=\epsilon\psi(x),
\end{equation}
where $\psi(x)=(\psi_1(x),i\psi_2(x))^T$ is a two-component spinor, see Eq.~(\ref{eq:1dDirac_free}).
The mass is set to be $m=939$~MeV/$c^2$, while we take a Woods-Saxon type for the potentials $V(x)$ and $S(x)$,
\begin{subequations}
\begin{eqnarray}
V(x)+S(x)&=&\frac{U_0}{1+e^{(|x|-R_U)/a_U}}\\
V(x)-S(x)&=&\frac{W_0}{1+e^{(|x|-R_W)/a_W}},
\end{eqnarray}
\end{subequations}
in which we use the parameters corresponding to $^{40}$Ca given in Ref.~\cite{Koepf1991}.

\begin{table}
\caption{The energy eigenvalues and the
expectation values of $p^2/2m$
for the bound eigenstates of a 1-dimensional Dirac equation without the
Wilson term.
These are obtained with the inverse Hamiltonian method with
the 3-point formula for the first derivative in the kinetic term. }
\begin{center}
\begin{tabular}{rrr}
\hline\hline
$k$ & $\epsilon_k$ (MeV) & $\langle p^2 \rangle/2m$ (MeV)\\
\hline
1 & $-65.8951$ & $   2.06~~~~~$ \\
2 & $-65.8951$ & $8291.43~~~~~$ \\
3 & $-52.1765$ & $   7.40~~~~~$ \\
4 & $-52.1765$ & $8286.09~~~~~$ \\
5 & $-34.4873$ & $  14.00~~~~~$ \\
6 & $-34.4873$ & $8279.48~~~~~$ \\
7 & $-16.6596$ & $  18.78~~~~~$ \\
8 & $-16.6596$ & $8274.71~~~~~$ \\
9 & $-2.9595$  & $  15.00~~~~~$ \\
10 & $-2.9595$ & $8278.49~~~~~$ \\
\hline\hline
\end{tabular}
\end{center}
\label{tb:energy-1d}
\end{table}

We discretize the coordinate $x$ with mesh size $a =0.1$~fm with $N=400$ mesh points and impose the box boundary condition.
We use the 3-point differential formula for the kinetic energy term.
The energy eigenvalues and expectation values of $p^2/2m$ for bound state solutions obtained with the inverse Hamiltonian method are summarized in Table~\ref{tb:energy-1d}.
We have checked that the dispersion of the Hamiltonian, $\langle h^2 \rangle-\langle h \rangle^2$, is close to zero for all the
states shown in the Table.
We find 5 pairs of bound states, and in each pair the two states have exactly the same energy.
The expectation value of $p^2$ is extremely large for one of the states in each pair, implying that it is a spurious solution.

\begin{figure}
\begin{center}
\includegraphics[angle=-90, width=8cm]{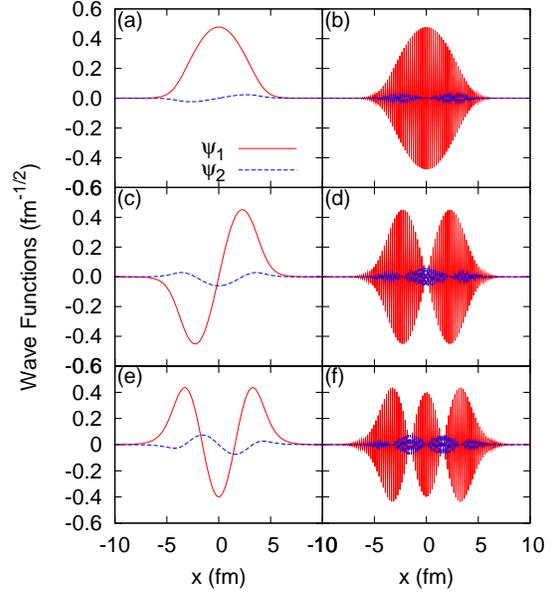}
\end{center}
\caption{(Color online) The wave functions for the six lowest energy eigenstates given in Table~\ref{tb:energy-1d}.
Panel (a) to (f) correspond to $k = 1$ to 6 in the Table, while the $\psi_1$ and $\psi_2$ correspond to the upper and lower components of the wave functions, respectively.}
\label{fig:wf-1d}
\end{figure}

In Fig.~\ref{fig:wf-1d}, we show the wave functions for the three lowest pairs.
The left and right panels correspond to the physical and the spurious solutions, respectively.
It is seen that the spurious states has the same amplitude as the corresponding physical states, but the sign of their wave functions alter at every mesh point.
More precisely, if we denote a physical state by $\psi^{(p)}(x_j)=\psi^{(p)}_j=(\psi_{1j}^{(p)},i\psi_{2j}^{(p)})^T$, and the corresponding spurious state by $\psi_j^{(s)}$, then it is given by
\begin{equation}
\psi_j^{(s)}=
\left(\begin{array}{c}
\psi_{1j}^{(s)}\\
i\psi_{2j}^{(s)}\\
\end{array}\right)
=
(-1)^j\left(\begin{array}{c}
\psi_{1j}^{(p)}\\
-i\psi_{2j}^{(p)}\\
\end{array}\right).
\end{equation}
It can be shown that, in the specific case of 3-point formula $f'_j=(f_{j+1}-f_{j-1})/2a$, the expectation values of kinetic term for the physical state and the corresponding spurious state are exactly the same.
Their expectation values of potential term are also the same.

\begin{figure}
\begin{center}
\includegraphics[angle=-90, width=8cm]{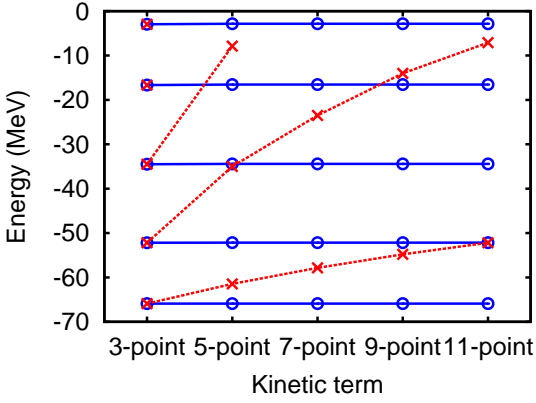}
\end{center}
\caption{(Color online) The
energy of the
bound states of the 1D Dirac equation obtained without the Wilson term.
The results are calculated with the 3-, 5-, 7-, 9-, and 11-point differential formulas for the kinetic term in the Hamiltonian.
The open circles and the crosses correspond to the physical and spurious states, respectively.
}
\label{fig:1dspe-der}
\end{figure}

We have shown that each pair of physical and spurious states is degenerate in energy if one uses the 3-point formula.
What happens to the spurious states if one approximates the derivative in the kinetic term with more accurate differential formula?
Figure~\ref{fig:1dspe-der} plots the spectra of the Dirac equation obtained with the 3-, 5-, 7-, 9-, and 11-point formulas for the kinetic term.
The physical and spurious solutions are shown by the open circles and the crosses, respectively.
They can be easily distinguished by monitoring the wave functions or the expectation values of $\langle p^2 \rangle/2m$.
The wave functions for an unphysical state oscillate very rapidly, and accordingly the expectation value of $\langle p^2 \rangle/2m$ is quite large compared to a typical value of the kinetic energy of a nucleon ($\lesssim 38$ MeV).
With the higher-order formulas, the spurious states are no longer degenerate to the corresponding physical states.
For instance, with the 5-point formula, the spurious states in the 4-th and 5-th pairs are already pushed away to the continuum region, while the lower spurious states are still in the bound region.
With the 7-point formula, the third
unphysical state also goes up to the continuum region.
The energy shift for the higher spurious states is larger than that for the lower spurious states.
Therefore, it is important that the higher-point differential formulas for the kinetic term resolve the degeneracy and lift up the spurious states, although the energy shift may not be large enough to remove all the spurious states from the energy region of interest.

\begin{figure}
\begin{center}
\includegraphics[angle=-90, width=8cm]{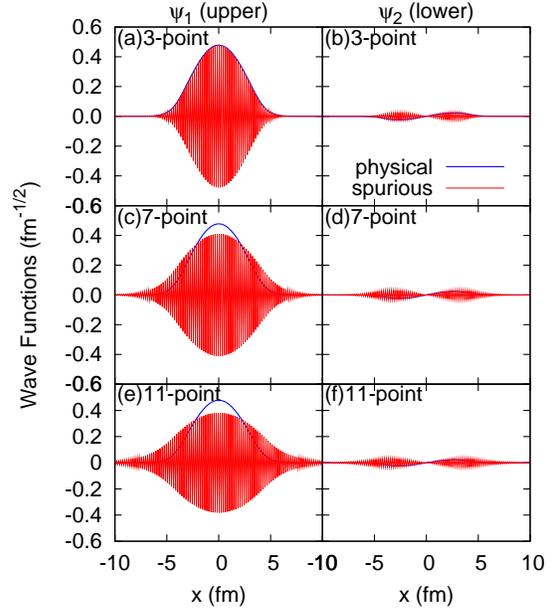}
\end{center}
\caption{(Color online) The wave functions for the lowest pair of states obtained without the Wilson term.
The results with
the 3-, 7-, and 11-point formulas are shown in the upper, middle, and lower panels, while the upper 
$\psi_1(x)$ and the lower $\psi_2(x)$ components of the wave functions are shown in the left and right panels, respectively.
}
\label{fig:wf-Nk}
\end{figure}

In Fig.~\ref{fig:wf-Nk}, we show the wave functions of the lowest pair of states obtained with the 3-, 7-, and 11-point formulas.
One can see that the amplitude of the spurious state deviates more from the physical state as the differential formula becomes more accurate.

\begin{figure}
\begin{center}
\includegraphics[width=8cm]{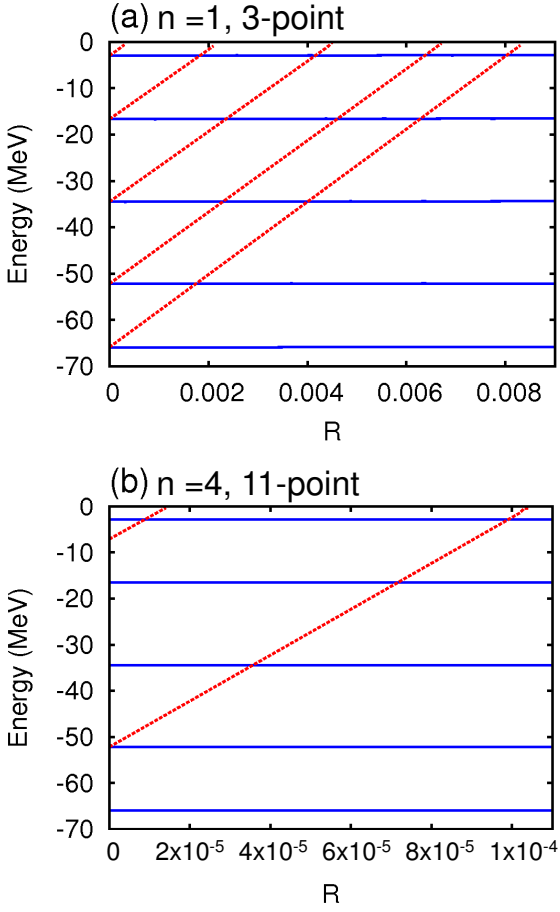}
\end{center}
\caption{(Color online) The energy spectra of the 1D Dirac equation as a function of the Wilson parameter $R$.
The solid and the dashed lines correspond to the physical and the spurious states, respectively.
The upper panel (a) shows the case with the original Wilson term ($n=1$) with the 3-point formula for both the kinetic and the Wilson terms.
The lower panel (b) shows the result with the 11-point formula for the kinetic term and an improved Wilson term ($n=4$) evaluated with the 9-point formula.
}
\label{fig:1dspe-wilp}
\end{figure}

Let us now switch on the Wilson term and investigate the change of the spectrum as a function of the Wilson parameter $R$.
In Fig.~\ref{fig:1dspe-wilp}, we show the energy spectrum of the Dirac equation as a function of $R$.
Figure~\ref{fig:1dspe-wilp}(a) is obtained with the original Wilson term ($n=1$).
The derivatives in the Wilson term and in the kinetic term are both approximated by the 3-point formula.
Figure~\ref{fig:1dspe-wilp}(b), on the other hand, shows the result with an high-order Wilson term ($n=4$)
evaluated with the 9-point formula, and the kinetic term is computed with the 11-point formula.
One can see that the energy shift for the unphysical states is proportional to the Wilson parameter $R$ in both cases.
In the case of $n=1$, all the unphysical states are pushed up to the continuum region around $R=0.008$, while all of them are pushed up to the continuum already at about $R=0.0001$ for $n=4$.
The high-order Wilson term is thus much more powerful than the original one.

\begin{figure}
\begin{center}
\includegraphics[angle=-90, width=8cm]{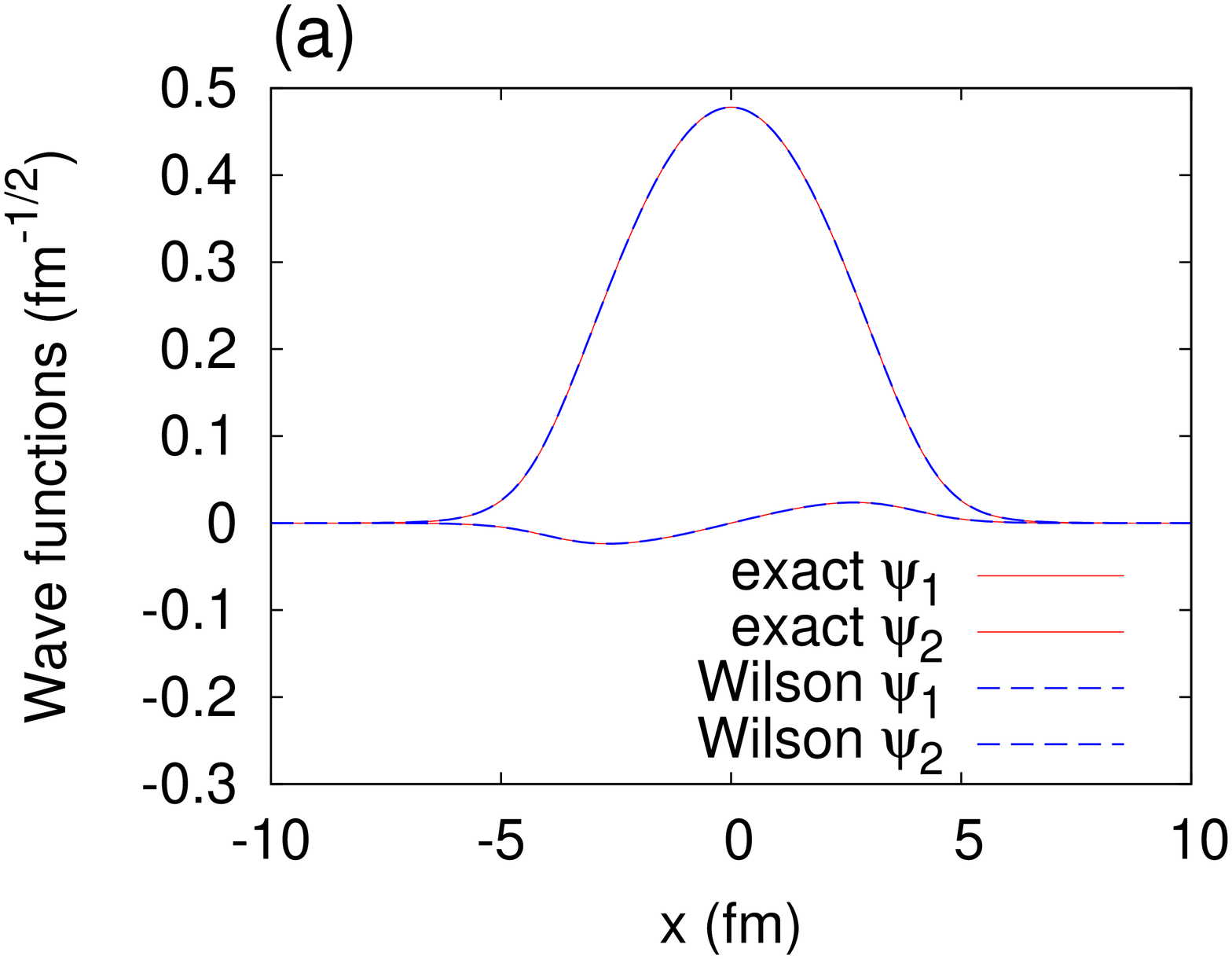}
\includegraphics[angle=-90, width=8cm]{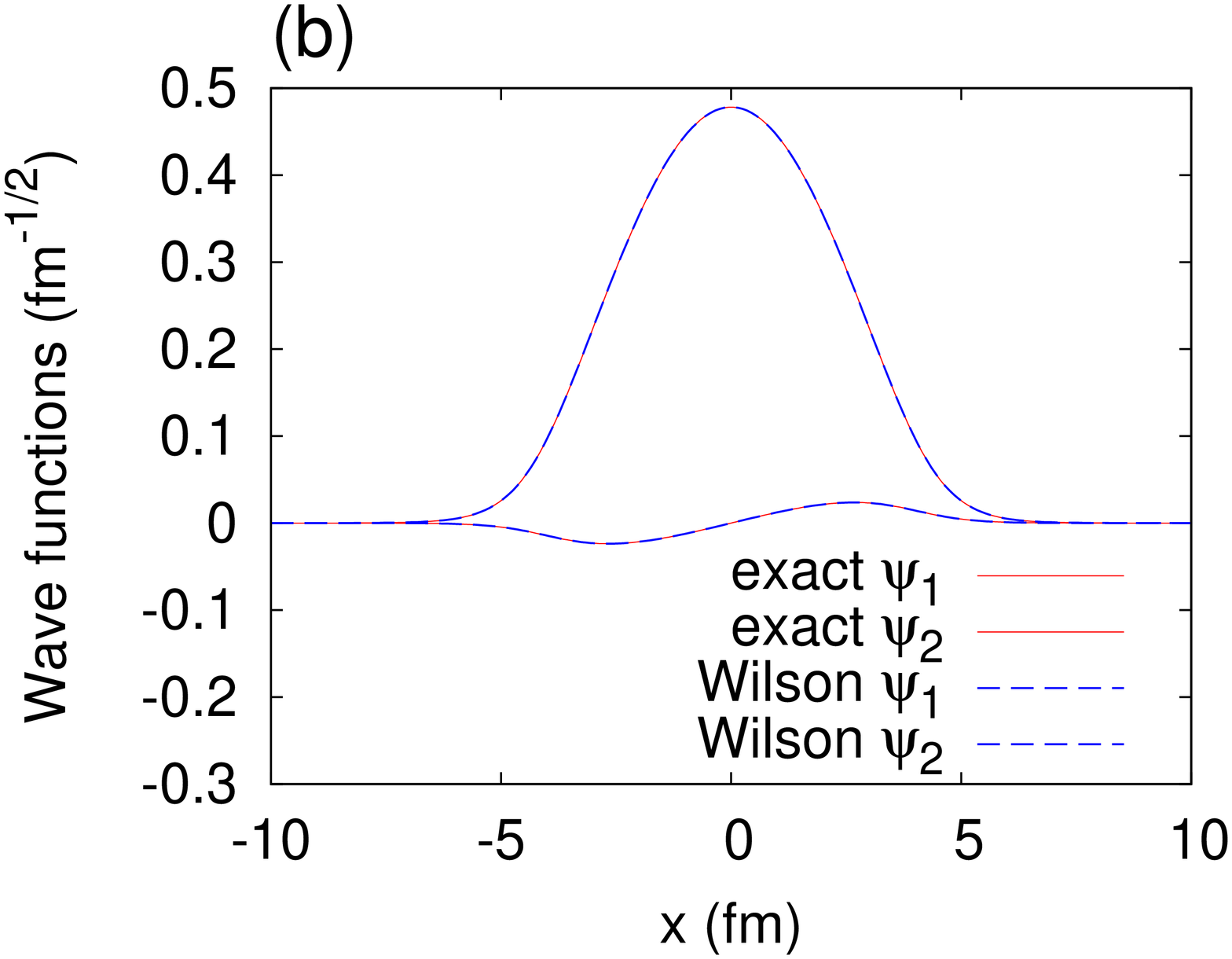}
\end{center}
\caption{(Color online)
A comparison of the wave functions obtained with the Wilson term (the dashed lines) to the ``exact'' ones (the solid lines) for the lowest single-particle state.
The exact wave function is obtained by solving the Dirac equation without the Wilson term by the inverse Hamiltonian method.
The upper panel (a) shows the case with the original Wilson term ($n=1$ and $R=0.01$) computed with the 3-point formula.
The lower panel (b) shows the result with a high-order Wilson term ($n=4$ and $R=0.00015$) computed with the 9-point formula.
In both cases the kinetic term is approximated by the 11-point formula.
The energy of this state with the Wilson term is $\epsilon=-65.8726$ MeV and $\epsilon=-65.8918$ MeV for the cases (a) and (b), respectively, while the exact energy is $\epsilon=-65.8918$ MeV. }
\label{fig:1dwf}
\end{figure}

Figure \ref{fig:1dwf} shows a comparison of the wave function for the lowest single-particle state obtained with the Wilson term to the exact one.
The upper panel shows the case with the original normal Wilson term ($n=1$ and $R=0.01$) with the 3-point formula.
On the other hand, the lower panel shows the result with an high-order Wilson term ($n=4$ and $R=0.00015$) computed by the 9-point formula.
In both cases, the kinetic term is approximated by the 11-point formula.
The ``exact'' wave function is obtained by solving the Dirac equation without the Wilson term by
the inverse Hamiltonian method. 
The energy of this state with the Wilson term after the corresponding correction is 
$\epsilon=\tilde{\epsilon}-\epsilon_W=-65.8726$~MeV and $\epsilon=-65.8918$~MeV for the cases 
(a) and (b), respectively, which can be compared with the exact energy $\epsilon=-65.8918$ MeV evaluated with the 11-point formula for the kinetic term.
Although the energies coincide between (a) and (b) only up to the first three digits, the difference in the wave functions is almost invisible in the scale shown in the figure.
In order to quantify the deviation in the wave functions, one can compute the overlap probability $|\langle \psi_{\rm ex}|\psi_{\rm W} \rangle|^2$, where $\psi_{\rm ex}$ is the exact wave function and $\psi_{\rm W}$ is the approximate wave function obtained with the Wilson term.
The deviation from unity is less than $10^{-14}$ for both cases.

In short, the properties of the Wilson fermion and a spectrum of spurious states are summarized as follows,
\begin{itemize}
\item Without the Wilson term, degenerate physical and unphysical states appear in pairs, if the first derivative in the kinetic term is approximated by the 3-point formula.
    When a more accurate differential formula is used in evaluating the kinetic term, the degeneracies are resolved, i.e., the spurious states are pushed upwards while the physical states stay unchanged.
    However, the energy shifts for the unphysical states are not always large enough to remove all the spurious states up to the continuum.
\item By switching on the Wilson term, all the spurious states are moved upwards by a similar amount of energy.
    The energy shifts are nearly proportional to the Wilson parameter $R$.
\item An increase of $n$ in the high-order Wilson term makes the shifts of the spurious states drastically large while the shifts of the physical states remain small.
    Thus a Wilson term with $n\geq 2$ has more effects on the spurious states and less effects on the physical states.
\item The solutions obtained with the high-order Wilson fermion are close to the exact solutions to a sufficient accuracy both in the energy eigenvalues and in the wave functions.
\end{itemize}

%

\end{document}